\documentclass{jfm}
\usepackage{graphicx}
\usepackage{epstopdf, epsfig}
\usepackage{amsmath}
\usepackage{amssymb,amsbsy,bm}
\usepackage{xcolor}
\usepackage[abs]{overpic}
\usepackage{subfiles}
\usepackage{cleveref}
\usepackage{subfig}
\crefformat{section}{\S#2#1#3} 
\crefformat{subsection}{\S#2#1#3}
\crefformat{subsubsection}{\S#2#1#3}
\usepackage{placeins}
\usepackage[normalem]{ulem}
\usepackage{tikz}
\usetikzlibrary{arrows.meta}
\usepackage{cancel}

\shorttitle{Multiscale preferential sweeping of particles settling in turbulence}
\shortauthor{J. Tom and A. D. Bragg}

\title{Multiscale preferential sweeping of particles settling in turbulence}

\author{Josin Tom\aff{1}
 \and Andrew D. Bragg\aff{1} \corresp{\email{andrew.bragg@duke.edu}}}

\affiliation{\aff{1} Department of Civil and Environmental Engineering, Duke University, Durham, NC, USA}

\begin{document}

\maketitle

\begin{abstract}
In a seminal article, \citet[J. Fluid Mech., 174:441-465]{maxey87} presented a theoretical analysis showing that enhanced particle settling speeds in turbulence occur through the preferential sweeping mechanism, which depends on the preferential sampling of the fluid velocity gradient field by the inertial particles. However, recent Direct Numerical Simulation (DNS) results in \citet[J. Fluid Mech., 796:659--711]{ireland16b} show that even in a portion of the parameter space where this preferential sampling is absent, the particles nevertheless exhibit enhanced settling velocities. Further, there are several outstanding questions concerning the role of different turbulent flow scales on the enhanced settling, and the role of the Taylor Reynolds number $R_\lambda$. The analysis of Maxey does not explain these issues, partly since it was restricted to particle Stokes numbers $St\ll1$. To address these issues, we have developed a new theoretical result, valid for arbitrary $St$, that reveals the multiscale nature of the mechanism generating the enhanced settling speeds. In particular, it shows how the range of scales at which the preferential sweeping mechanism operates depends on $St$. This analysis is complemented by results from DNS where we examine the role of different flow scales on the particle settling speeds by coarse-graining the underlying flow. The results show how the flow scales that contribute to the enhanced settling depend on $St$, and that contrary to previous claims, there can be no single turbulent velocity scale that characterizes the enhanced settling speed. The results explain the dependence of the particle settling speeds on $R_\lambda$, and show how the saturation of this dependence at sufficiently large $R_\lambda$ depends upon $St$. The results also show that as the Stokes settling velocity of the particles is increased, the flow scales of the turbulence responsible for enhancing the particle settling speed become larger. Finally, we explored the multiscale nature of the preferential sweeping mechanism by considering how particles preferentially sample the fluid velocity gradients coarse-grained at various scales. The results show that while rapidly settling particles do not preferentially sample the fluid velocity gradients, they do preferentially sample the fluid velocity gradients coarse-grained at scales outside of the dissipation range. This explains the findings of Ireland \emph{et al.}, and further illustrates the truly multiscale nature of the mechanism generating enhanced particle settling speeds in turbulence.

\end{abstract}

\begin{keywords}
\end{keywords}

\section{Introduction}\label{sec:Intro}
The motion of inertial particles in turbulence settling under gravity is important for many environmental, biological and engineering multi-phase flows such as water droplets in clouds \citep{shaw03, grabowski13}, marine snow \citep{kiorboe97,guseva16} and sediment transport \citep{papanicolaou08}. For these problems, it is important to quantify the settling speed of the particles as they fall through the turbulent flow, since this determines the vertical mass flux of the particles. In a still fluid of infinite extent, small particles settle with the Stokes settling velocity \citep{batchelor67}, which is determined by the balance of drag and gravity forces acting on the particles. In a turbulent flow, the drag forces acting on the particles fluctuate, and an important question is whether these fluctuations modify the average settling velocity of the particles compared with that in a still fluid. 

The answer to this question depends upon the parameters of the system. For the case of small (diameter much smaller than the Kolmogorov length scale $\eta$), and heavy (particle density much greater than fluid density) inertial particles, two important parameters are the particle Stokes number, $St$, and the Froude number, $Fr$. The Stokes number, $St\equiv \tau_p/\tau_\eta$, provides a measure of the particle inertia, where $\tau_p$ is the particle response time and $\tau_\eta$ is the Kolmogorov timescale. The Froude number, $Fr \equiv a_{\eta}/g$, quantifies the strength of the turbulence relative to gravity, where $a_{\eta}$ is the Kolmogorov acceleration scale and $g$ is the magnitude of the gravitational acceleration. From $St$ and $Fr$ the settling parameter can be defined $Sv \equiv St/Fr\equiv \tau_p g/u_\eta$, that compares the Stokes settling velocity to the Kolmogorov velocity scale $u_\eta$. Another important parameter is the particle Reynolds number, $Re_{p}$, that determines whether the drag force on the particle is a linear or non-linear function of the slip velocity between the particle and local fluid velocity \citep{maxey83}. It is typically assumed that the drag force is linear if $Re_p<0.5$ \citep{elghobashi93}.

For particles subject to a linear drag force, \citet{reeks77} argued that in a stochastic random fluid velocity field, there would be no net effect of the flow fluctuations on the average particle settling velocity. 
In contrast, \citet{maxey86} performed a numerical study and demonstrated that aerosol particles subject to a linear drag force settle more rapidly in a randomly oriented, periodic, cellular flow field than they do in a still fluid. In a seminal article, \citet{maxey87} reported enhanced particle settling velocities in simulations using a linear drag force and a Gaussian fluid velocity field. He then presented a detailed theoretical analysis for $St\ll 1$ to show that enhanced particle settling speeds in spatio-temporally correlated fluid velocity fields can occur because inertial particles preferentially sample low vorticity regions of the fluid velocity field where the flow is moving down (in the direction of gravity). Maxey argued that this preferential sampling occurs because inertial particles are centrifuged out of regions of strong vorticity, something that was subsequently confirmed by \citet{squires91a} using Direct Numerical Simulations (DNS) of particle motion in turbulence. In a subsequent work, \citet{wang93} performed DNS for small, heavy settling particles subject to linear drag, and observed significant enhancement of the particle settling velocities due to turbulence for particles with $Sv=O(1)$ and $St=O(1)$. They also confirmed the physical argument of \citet{maxey87} that settling inertial particles preferentially accumulate in low vorticity regions of the flow where fluid velocity aligns with the direction of gravity. They referred to this effect as the ``preferential sweeping mechanism'' or ``fast-tracking'', and it has since been observed in several DNS \citep{bec14b,ireland16b,rosa16,monchaux17} and experimental studies \citep{aliseda02,good14,petersen19}.

 When $Re_p>1$, a non-linear drag law should be used for calculating the drag force on the particle \citep{clift78}. 
 Early experimental works \citep{tunstall67,schoneborn75} have shown that the combination of particle inertia and non-linear drag can reduce the settling velocity in simple oscillatory flows.  
 Two common ways of introducing non-linear drag are to use a simple square drag law \citep{fung93}, or use an empirical relationship for the drag coefficient based on the instantaneous particle Reynolds number \citep{fung98}. \citet{fung93} used a simple Gaussian fluid velocity field with a square drag law, and observed a decrease of the particle settling velocities compared with the linear drag case, as did \citet{mei94} who prescribed the fluid velocity using a stochastic process and employed a non-linear drag coefficient. In contrast, \citet{chan99} found an increase in the particle settling velocity due to non-linear drag for particles settling in a simple, two-dimensional periodic cellular flow with a square drag law. \citet{nielsen84} used an analytical solution approach to argue that the reduction in the particle settling velocity due to non-linear drag is negligible. 
 
 These studies on the effects of non-linear drag on particle settling speeds did not, however, consider real turbulence, but only simple random flow fields. Nevertheless, DNS studies have also come to conflicting conclusions concerning the role of non-linear drag on particle settling speeds in turbulence.  \citet{wang93} studied the influence of a non-linear drag coefficient using DNS and found that its effect was to only slightly reduce the particle settling speeds compared to the linear drag case, with similar results in \citet{rosa16}. However, \citet{good14} observed a significant reduction in the particle settling velocity compared with the linear drag case, in certain regimes of $St, Sv$.
 
 Another issue is whether in some cases it is possible for turbulence to reduce the particle settling speed compared with the Stokes settling velocity. This scenario, referred to by \citet{nielsen93} as ``loitering'', could occur when the particles spend more time in upward moving regions of the flow than downward moving regions. In their DNS study using non-linear drag, \citet{wang93} observed loitering only in a limited portion of the parameter space.  Loitering has been observed in several experiments \citep{yang03, kawanisi08,good14}, while the recent detailed measurements of \cite{petersen19} did not show clear evidence of loitering. \citet{good14} were only able to observe loitering in their DNS when either the horizontal motion of the particles was artificially eliminated, or else when a non-linear drag force was used for the particles. However, as discussed earlier, other studies such as \citet{rosa16} have found only a small effect of non-linear drag on the particle settling speeds, and they did not observe loitering.

 These studies show that even for the relatively simple case of small particles settling in turbulence, the role of turbulence on the average particle settling speed is subtle, and a number of issues remain to be solved. Moreover, there are other additional complexities that can modify particle settling speeds in turbulence, including finite particle size effects \citep{fornari16}, particle-fluid two-way coupling \citep{monchaux17} and collective particle interaction effects \citep{huck18}, all of which significantly complicate the problem.
 
 Even in the absence of these additional complexities, an aspect of the problem that has not been systematically explored concerns the role of different turbulent flow scales in modifying the particle settling speeds in turbulence. Due in part to its restriction to $St\ll1$, the theoretical analysis of \citet{maxey87} gives little insight into which scales of the turbulence contribute to the particle settling speeds. Numerical studies, have, to a limited extent, considered the question of which flow scales contribute to the enhanced settling. \cite{wang93} argued that their DNS results implied that the enhanced particle settling speeds due to turbulence depends on the fluid r.m.s. velocity $u'$, which is associated with the large scales of the turbulent flow. However, they also argued that this could be an artifact of the low Reynolds number of their DNS, and that in real atmospheric flows where the scale separation is much larger, the particle settling speeds are likely to be only affected by a limited range of scales of the turbulence. The study of \cite{yang98} considered the role played by different flow scales on the particle settling speeds in turbulence using DNS and Large Eddy Simulations (LES). They concluded that the large scales of the flow play a key role and that the relevant fluid velocity scale determining the settling enhancement is $u'$ and not $u_\eta$. However, they also concluded that the settling enhancement depends on the particle inertia through $\tau_p/\tau_\eta$, and not $\tau_p/\tau_L$, where $\tau_L$ is the integral timescale of the flow. Their argument is that the preferential sweeping effect depends on the small scale clustering of the particles, and hence $\tau_\eta$, while the drag force on the particles depends mainly on the large scales, and hence $u'$. The recent study of \cite{rosa16} also argued that the relevant fluid velocity scale determining the settling enhancement is $u'$. 

 The success of ``mixed scaling'', i.e. using a large scale quantity $u'$ for the velocity scale, and a small scale quantity $\tau_\eta$ for the time scale, used in previous studies \citep{yang98,good14}, reveal something of the multiscale nature of the problem, and that both large and small scales of the turbulence affect the settling speeds. However, there is a need to elucidate whether the single velocity scale $u'$ really does dominate the particle settling speeds, or whether there exists a range of velocity scales that govern the process. Indeed, as we shall point out in \S\ref{sec:Theory}, there are theoretical difficulties with the argument that the relevant turbulent velocity scale dominating the settling speed is $u'$. Moreover, it is not at all obvious that a single velocity scale should determine the settling speed, independent of $St$, since the interaction of an inertial particle with a given flow scale depends strongly on $St$. Motivated by the need to develop particle SubGrid Scale (SGS) models for LES of particles settling in turbulent flows, \cite{rosa17} used numerical simulations to study the effect of the small scales of the turbulence on particle settling speeds. Their results showed that scales smaller than a certain size did not affect the particle settling speeds. However, they did not consider whether there exists an upper limit to the range of scales that affect the particle settling speeds, nor did they provide detailed information concerning how the range of velocity scales impacting the settling speeds might depend on $St$ or $Fr$.

 In addition to these open issues, results in \cite{ireland16b} showed that significant enhancement of particle settling speeds due to turbulence are observed for $St\geq O(1)$ and $Fr\ll 1$, even though the DNS data showed that for $St\geq O(1)$ and $Fr\ll 1$ the particles do not preferentially sample the fluid velocity gradient field. This seems inconsistent with the preferential sweeping mechanism which relies on the idea that inertial particles do preferentially sample the fluid velocity gradient field, showing a preference to avoid regions dominated by vorticity due to the centrifuge effect. 

 In order to resolve these issues and provide insight into the role of different flow scales on the enhanced settling speeds of particles in turbulence, it is desirable to develop a theoretical framework that goes beyond the work of \citet{maxey87} that restricted attention to $St\ll 1$. The purpose of the present study then is to address these issues and provide insight concerning which flow scales contribute to the modified particle settling speeds due to turbulence. To this end, we develop a new theoretical framework for analyzing the problem for arbitrary $St$, which is then used in conjunction with DNS data for a range of Reynolds numbers to provide detailed insights into the multiscale mechanism leading to enhanced particle settling speeds in turbulence.
\section{Theory}\label{sec:Theory}
\subsection{Background}
We consider the settling of small ($d_p/\eta \ll 1$, where $d_p$ is the particle diameter), heavy ($\rho_p/\rho_f \gg 1$, where $\rho_p$ is particle density and $\rho_f$ is fluid density), spherical inertial particles, that are one-way coupled to a statistically stationary isotropic turbulent flow. In the regime of a linear drag force on the particles, the particle equation of motion reduces to \citep[see][]{maxey83} 
\begin{align}
\ddot{\bm{x}}^p(t)\equiv\dot{\bm{v}}^p(t)=\frac{1}{St\tau_\eta}\Big(\bm{u}(\bm{x}^p(t),t)-\bm{v}^p(t)\Big)+\bm{g},\label{eom}
\end{align}
where $\bm{x}^p(t),\bm{v}^p(t)$ are the particle position and velocity vectors, respectively, $\bm{u}(\bm{x}^p(t),t)$ is the fluid velocity at the particle position, and $\bm{g}$ is the gravitational acceleration vector.

As discussed earlier, some studies suggest that for $Sv\geq O(10)$, nonlinear drag effects are important for settling particle motion in turbulence \citep{good14}, while other studies suggest that even up to $Sv\approx 63$, the effects of nonlinear drag are very small \citep{rosa16}. Such discrepancies must be resolved in future work. However, in the present study we will ignore nonlinear corrections to the drag force for analytical simplicity, leaving the consideration of nonlinear drag, as well as other complexities such as two-way momentum coupling with the fluid, for future work.

For the system described above, the ensemble average of \eqref{eom} in the direction of gravity $\bm{e}_z$ gives 
\begin{align}
\langle{v}_z^p(t)\rangle=\langle{u}_z(\bm{x}^p(t),t)\rangle+St\tau_\eta{g},\label{SV}
\end{align}
since $\langle \dot{\bm{v}}^p(t)\rangle=\bm{0}$ for this system. Equation \eqref{SV} shows that the average particle velocity may differ from the Stokes settling velocity $St\tau_\eta{g}$ only if $\langle{u}_z(\bm{x}^p(t),t)\rangle\neq{0}$. Numerous studies, both numerical \citep{wang93,bec14b,ireland16b,rosa16,monchaux17} and experimental \citep{aliseda02,good14,petersen19}, have indeed shown that $\langle {v}_z^p(t)\rangle\neq St\tau_\eta{g}$, implying $\langle{u}_z(\bm{x}^p(t),t)\rangle\neq{0}$. \citet{maxey87} developed a theoretical framework to explain how $\langle{u}_z(\bm{x}^p(t),t)\rangle\neq{0}$, even though the Eulerian average satisfies $\langle{u}_z(\bm{x},t)\rangle={0}$ for an isotropic flow. Essentially, the explanation is that particles with inertia do not uniformly sample the underlying fluid velocity field, and that gravity leads to a bias for inertial particles to accumulate in regions of the flow where $\bm{e}_z\bm{\cdot u}>0$. However, the analysis of \citet{maxey87} is restricted to $St\ll1$.

A key question concerns which scales of the turbulent flow influence the quantity $\langle{u}_z(\bm{x}^p(t),t)\rangle$. While this has not previously been systematically explored, one claim that has been made in numerous studies \citep{wang93,yang98,good14,nemes17,petersen19} is that $\langle{u}_z(\bm{x}^p(t),t)\rangle$ depends on $u'$, the large scale fluid velocity. However, this seems unlikely because if $\langle{u}_z(\bm{x}^p(t),t)\rangle$ depends on $u'$, then even if $Sv=O(1)$, $\langle{u}_z(\bm{x}^p(t),t)\rangle/u_\eta \to\infty$ as $R_\lambda\to\infty$ since $u'/u_\eta\sim R_\lambda^{1/2}$ \citep{pope}. Yet this seems to lead to a contradiction since
\begin{align}
\langle{v}_z^p(t)\rangle/u_\eta=\langle{u}_z(\bm{x}^p(t),t)\rangle/u_\eta+Sv,\label{SV2}
\end{align}
so that for $Sv=O(1)$, the particle settling becomes irrelevant to the mean motion of the particle in the limit $R_\lambda\to\infty$, yet the settling is supposed to be the very thing responsible for $\langle{u}_z(\bm{x}^p(t),t)\rangle\neq 0$. Clearly then new insight is needed to understand which scales of the turbulent flow influence $\langle{u}_z(\bm{x}^p(t),t)\rangle$, and we now develop a new theoretical framework to provide such insight.
\subsection{Theoretical framework for arbitrary $St$}\label{TFAS}
In this section we develop a theoretical framework for considering the behavior of $\langle{u}_z(\bm{x}^p(t),t)\rangle$ for arbitrary $St$, and for revealing which scales of motion contribute to $\langle{u}_z(\bm{x}^p(t),t)\rangle\neq 0$.

In the analysis of \cite{maxey87}, attention was restricted to $St\ll1$, for which it is possible to approximate $\bm{v}^p(t)$ as a field
\begin{align}
\begin{split}
\bm{v}^p(t)=\Big(\bm{u}(\bm{x},t)-St\tau_\eta[\bm{a}(\bm{x},t)-\bm{g}]\Big)\Big\vert_{\bm{x}=\bm{x}^p(t)}+O(St^2),\label{Mpvv}
\end{split}
\end{align}
where $\bm{x}$ is a fixed point in space (unlike $\bm{x}^p(t)$), and $\bm{a}\equiv \partial_t\bm{u}+(\bm{u\cdot\nabla})\bm{u}$ is the fluid acceleration field. Using this field representation, Maxey was able to construct an expression for $\langle{u}_z(\bm{x}^p(t),t)\rangle$ using a continuity equation for the instantaneous particle number density. However, when $St\geq O(1)$, this field approximation for $\bm{v}^p(t)$ fundamentally breaks down due to the formation of caustics in the particle velocity distributions, wherein particle velocities at a given location become multivalued \citep{wilkinson05}. Therefore, a quite different approach to that employed by \cite{maxey87} must be sought in order to analyze $\langle{u}_z(\bm{x}^p(t),t)\rangle$ for arbitrary $St$.

We begin by noting that for homogeneous turbulence we may write
\begin{align}
\langle{u}_z(\bm{x}^p(t),t)\rangle=\varrho^{-1}\Big\langle{u}_z(\bm{x},t)\delta(\bm{x}^p(t)-\bm{x})\Big\rangle,\label{uav2}
\end{align}
where $\delta(\cdot)$ is a Dirac distribution, and $\varrho\equiv\langle \delta(\bm{x}^p(t)-\bm{x})\rangle$ is the Probability Density Function (PDF) of $\bm{x}^p(t)$. Here, $\langle\cdot\rangle$ is an ensemble average over all possible realizations of the system, and this includes not only an average over all realizations of $\bm{u}$, but also an average over all initial particle positions $\bm{x}^p(0)=\bm{x}_0$ and velocities $\bm{v}^p(0)=\bm{v}_0$. From \eqref{uav2} it follows that for a homogeneous turbulent flow, $\langle{u}_z(\bm{x}^p(t),t)\rangle=\langle{u}_z(\bm{x},t)\rangle=0$ if $\bm{x}^p(t)$ is uncorrelated with $u_z(\bm{x},t)$. This occurs for $St\gg 1$ particles that move ballistically. It also occurs for fluid particles that are fully-mixed at $t=0$ \citep{bragg12}, since their spatial distribution remains constant and uniform $\forall t$ due to incompressibility. In fact, $\langle{u}_z(\bm{x}^p(t),t)\rangle\neq{0}$ can only occur if $\delta(\bm{x}^p(t)-\bm{x})$ both fluctuates in time and is also correlated with $u_z(\bm{x},t)$.

To proceed with an analysis of $\langle{u}_z(\bm{x}^p(t),t)\rangle$ that applies for arbitrary $St$, we introduce the averaging decomposition $\langle\cdot\rangle=\langle\langle\cdot\rangle^{\bm{x}_0,\bm{v}_0}_{\bm{u}}\rangle^{\bm{u}}$ \citep{bragg12b}, where $\langle\cdot\rangle^{\bm{x}_0,\bm{v}_0}_{\bm{u}}$ denotes an average over all initial particle positions $\bm{x}^p(0)=\bm{x}_0$ and velocities $\bm{v}^p(0)=\bm{v}_0$ for a given realization of the fluid velocity field $\bm{u}$, and $\langle\cdot\rangle^{\bm{u}}$ denotes an average over all realizations of $\bm{u}$. Introducing this decomposition to \eqref{uav2} we have
\begin{align}
\langle{u}_z(\bm{x}^p(t),t)\rangle=\varrho^{-1}\Big\langle\Big\langle{u}_z(\bm{x},t)\delta(\bm{x}^p(t)-\bm{x})\Big\rangle^{\bm{x}_0,\bm{v}_0}_{\bm{u}}\Big\rangle^{\bm{u}}=\varrho^{-1}\Big\langle{u}_z(\bm{x},t)\varphi(\bm{x},t)\Big\rangle^{\bm{u}},\label{uavD}
\end{align}
where 
\begin{align}
\varphi(\bm{x},t)\equiv \Big\langle\delta(\bm{x}^p(t)-\bm{x})\Big\rangle^{\bm{x}_0,\bm{v}_0}_{\bm{u}},
\end{align}
and also $\varrho\equiv\langle\varphi(\bm{x},t)\rangle^{\bm{u}}$. The reason for introducing this averaging decomposition is that it will allow us to introduce a particle velocity field that is valid for arbitrary $St$, unlike \eqref{Mpvv} that is only valid for $St\ll1$.

The evolution equation for $\varphi$ is given by 
\begin{align}
\partial_t\varphi+ \bm{\nabla\cdot}\Big(\varphi\bm{\mathcal{V}}(\bm{x},t)\Big)=0,\label{phieq}
\end{align}
where
\begin{align}
\bm{\mathcal{V}}(\bm{x},t)&\equiv \Big\langle\bm{v}^p(t)\Big\rangle^{\bm{x}_0,\bm{v}_{0}}_{\bm{x},\bm{u}}.
\end{align}
The particle velocity field $\bm{\mathcal{V}}(\bm{x},t)$ differs fundamentally from the particle velocity field used in \cite{maxey87}, namely \eqref{Mpvv}, since $\bm{\mathcal{V}}(\bm{x},t)$ does not presume that $\bm{v}^p(t)$ is uniquely determined for $\bm{x}^p(t)=\bm{x}$ in a given realization of $\bm{u}$. Rather, $\bm{\mathcal{V}}(\bm{x},t)$ is constructed as an average over different particle trajectories (each corresponding to different $\bm{x}_0,\bm{v}_0$) satisfying $\bm{x}^p(t)=\bm{x}$ in a given realization of $\bm{u}$. We also emphasize that both $\varphi$ and $\bm{\mathcal{V}}$ are turbulent fields, in general, since they depend upon the evolution of the particular realization of $\bm{u}$ to which they correspond.

The solution to \eqref{phieq} may be written formally as
\begin{align}
\varphi(\bm{x},t)=\varphi(\bm{\mathcal{X}}(0\vert\bm{x},t),0)\exp\Bigg( -\int_0^t \bm{\nabla\cdot}\bm{\mathcal{V}}(\bm{\mathcal{X}}(s\vert\bm{x},t),s)\,ds \Bigg),\label{phisol}
\end{align}
where $\dot{\bm{\mathcal{X}}}(t)\equiv \bm{\mathcal{V}}(\bm{\mathcal{X}}(t),t)$, and the notation $s\vert\bm{x},t$ denotes that the variable is measured at time $s$ along a trajectory satisfying $\bm{\mathcal{X}}(t)=\bm{x}$. Note that $ \bm{\nabla\cdot}\bm{\mathcal{V}}(\bm{\mathcal{X}}(s\vert\bm{x},t),s)$ is to be understood as\[\bm{\nabla\cdot}\bm{\mathcal{V}}(\bm{\mathcal{X}}(s\vert\bm{x},t),s)\equiv \bm{\nabla\cdot}\bm{\mathcal{V}}(\bm{y},s)\Bigg\vert_{\bm{y}=\bm{\mathcal{X}}(s\vert\bm{x},t)}\] such that the operator $\bm{\nabla\cdot}\{\}$ acts on the spatial coordinate of the field $\bm{\mathcal{V}}$ and not on the trajectory end-point coordinate $\bm{x}$.

For simplicity, we will take $\varphi(\bm{\mathcal{X}}(0\vert\bm{x},t),0)=1/\mathcal{D}$, corresponding to particles that are initially uniformly distributed throughout the volume $\mathcal{D}$ of the system (this was also assumed in \cite{maxey87}). Further, we note that for a statistically stationary, homogeneous system, $\varrho\mathcal{D}=1$ $\forall t$ when $\varphi(\bm{\mathcal{X}}(0\vert\bm{x},t),0)=1/\mathcal{D}$. Using this initial condition for all realizations of $\bm{u}$, we may then insert \eqref{phisol} into \eqref{uavD} and obtain
\begin{align}
\langle{u}_z(\bm{x}^p(t),t)\rangle=\Bigg\langle{u}_z(\bm{x},t)\exp\Bigg( -\int_0^t \bm{\nabla\cdot}\bm{\mathcal{V}}(\bm{\mathcal{X}}(s\vert\bm{x},t),s)\,ds \Bigg)\Bigg\rangle^{\bm{u}}.\label{GR2}
\end{align}
%
 
Before proceeding, we note that in the regime $St\ll1$ we may insert \eqref{Mpvv} into the definition of $\bm{\mathcal{V}}(\bm{x},t)$ and obtain
\begin{align}
\begin{split}
\bm{\mathcal{V}}(\bm{x},t)&= \Big\langle  \bm{u}(\bm{x}^p(t),t)-St\tau_\eta[\bm{a}(\bm{x}^p(t),t)-\bm{g}] \Big\rangle^{\bm{x}_0,\bm{v}_{0}}_{\bm{x},\bm{u}}+O(St^2)\\
&=\bm{u}(\bm{x},t)-St\tau_\eta[\bm{a}(\bm{x},t)-\bm{g}]+O(St^2),
\end{split}
\end{align}
and inserting this into \eqref{GR2} we obtain essentially the same result as \cite{maxey87}
\begin{align}
\langle{u}_z(\bm{x}^p(t),t)\rangle\approx\Bigg\langle{u}_z(\bm{x},t)\exp\Bigg( St\tau_\eta\int_0^t {\Big(} \mathcal{S}^2(\bm{x}^p(s\vert\bm{x},t),s)-\mathcal{R}^2(\bm{x}^p(s\vert\bm{x},t),s) {\Big)}\,ds \Bigg)\Bigg\rangle^{\bm{u}},\label{GR3}
\end{align}
where $\mathcal{S}^2$ and $\mathcal{R}^2$ are the second invariants of the fluid strain-rate and rotation-rate tensors, respectively. The interpretation of \eqref{GR3} given by \cite{maxey87} is that $\langle{u}_z(\bm{x}^p(t),t)\rangle>0$ arises because settling inertial particles preferentially sweep around vortices where $\mathcal{S}^2-\mathcal{R}^2>0$ due to the centrifuging effect, favoring the downward moving side of the vortices associated with ${u}_z>0$.

Our result in \eqref{GR2}, that is not restricted to $St\ll1$, suggests more generally that $\langle{u}_z(\bm{x}^p(t),t)\rangle>0$ can arise when $\bm{\mathcal{V}}(\bm{x},t)$ is compressible, and also when there exists a correlation between regions where $\bm{\nabla\cdot}\bm{\mathcal{V}}<0$ and ${u}_z>0$. The argument given by \cite{wang93} for this correlation is essentially connected to the fact that settling particles typically approach the turbulent vortices from above as they fall through the flow, and they are swept around the vortices due to the centrifuge effect. A supplementary argument to theirs is that the settling particles tend to follow the ``path of least resistance''. In particular, downward moving particles prefer to move around the downward moving side of vortices in the flow since on this side they experience a weaker drag force (``less resistance'') than they would if they were to fall around the upward moving side of the vortice.

Unlike the $St\ll 1$ regime, for $St\geq O(1)$, $\bm{\mathcal{V}}(\bm{x},t)$ depends non-locally in time upon the fluid velocity field. Indeed, using the formal solution to \eqref{eom} we may write (ignoring initial conditions)
\begin{align}
\begin{split}
\bm{\mathcal{V}}(\bm{x},t)=St\tau_\eta\bm{g}+\frac{1}{St\tau_\eta}\int_0^t e^{-(t-s)/St\tau_\eta}\Big\langle  \bm{u}(\bm{x}^p(s),s)\Big\rangle^{\bm{x}_0,\bm{v}_{0}}_{\bm{x},\bm{u}}\,ds,
\end{split}
\end{align}
so that $\bm{\mathcal{V}}(\bm{x},t)$ depends on $\bm{u}$ at earlier times along the particle trajectory, and is in this sense temporally non-local. Due to this non-locality, there need not exist a correlation between $\bm{\nabla\cdot}\bm{\mathcal{V}}$ and the local properties of the flow. When $\bm{\nabla\cdot}\bm{\mathcal{V}}$ is uncorrelated with the fluid velocity field then from \eqref{GR2} we have
\begin{align}
\begin{split}
\langle{u}_z(\bm{x}^p(t),t)\rangle&=\Bigg\langle{u}_z(\bm{x},t)\exp\Bigg( -\int_0^t \bm{\nabla\cdot}\bm{\mathcal{V}}(\bm{\mathcal{X}}(s\vert\bm{x},t),s)\,ds \Bigg)\Bigg\rangle^{\bm{u}}\\
&=\Bigg\langle{u}_z(\bm{x},t)\Bigg\rangle^{\bm{u}}\Bigg\langle \exp\Bigg( -\int_0^t \bm{\nabla\cdot}\bm{\mathcal{V}}(\bm{\mathcal{X}}(s\vert\bm{x},t),s)\,ds \Bigg)\Bigg\rangle^{\bm{u}}\\
&=0.
\end{split}
\end{align}
Thus it is not merely clustering of the particles (related to $\bm{\nabla\cdot}\bm{\mathcal{V}}<0$) that is required for $\langle{u}_z(\bm{x}^p(t),t)\rangle>0$, but that the clustering be correlated in some way with the fluid velocity field. In view of this it is essential to make a distinction between two phenomena, namely particle clustering and preferential concentration. As emphasized in \cite{bragg14d}, these are distinct: clustering refers to non-uniformity of the particle spatial distribution, irrespective of any correlation the distribution may have with the fluid flow field. In contrast, preferential concentration describes the situation where the spatial distribution of the particles is not only non-uniform, but is also correlated to the local properties of the flow i.e. the particles cluster in specific regions of the flow, so that the particles preferentially sample the flow field. Recent results have shown that for settling particles this distinction is particularly important, since settling inertial particles can strongly cluster in the dissipation range of turbulence, despite the fact that their spatial distribution is entirely uncorrelated with the dissipation range properties of the turbulence \citep{ireland16b}. Indeed, while \cite{ireland16b} showed that the particle clustering (measured by the Radial Distribution Function) becomes monotonically stronger at progressively smaller scales for all $St$, in \S\ref{pref_samp} we will show that the preferential sampling of the turbulent flow field is strongest at some intermediate flow scale that depends on $St$. This can occur because when $St\geq O(1)$, the mechanism that causes the particle clustering is not the centrifuge mechanism and the associated preferential sampling of the flow field discussed in \cite{maxey87}, but rather a non-local mechanism that does not depend upon the particle interaction with the local fluid velocity field \citep{bragg14b,bragg14d,bragg14e,gustavsson11b}.

A subtle but important point that follows from this is that it is not in general appropriate to test the validity of the preferential sweeping mechanism by considering the particle settling velocities conditioned on the local particle concentration, as has often been done \citep{wang93,rosa17,petersen19}. The reason is that, as pointed out above, it is preferential concentration, and not clustering (which is measured by local particle concentrations), that is connected with preferential sweeping and the enhanced particle settling speeds. The driver of the preferential sweeping mechanism is not the strength of the clustering (quantified by the local concentration), but the degree to which the particle locations are correlated with the local flow, i.e. preferential concentration. However, as discussed above, when $St\geq O(1)$, the clustering of settling particles may not be correlated with the local flow, and therefore analyzing the particle settling velocities conditioned on the local particle concentration no longer provides a meaningful test of the validity of the preferential sweeping mechanism. It is interesting to note that \cite{rosa17} observed that particle settling velocities increase with increasing local particle concentration for low inertia particles, but an opposite and weaker trend was observed for high inertia particles. This reversal in trend was attributed to the loitering effect and the ineffectiveness of the preferential sweeping mechanism for higher inertia particles. While this interpretation in terms of loitering may be valid, the differing behavior observed for low and high $St$ may be simply a reflection of the fact that while the settling velocity conditioned on concentration is an appropriate measure of settling velocity enhancement for $St \ll 1$, it is not for $St\geq O(1)$, as explained before.

\subsection{Multiscale insight}
Having constructed an expression for $\langle{u}_z(\bm{x}^p(t),t)\rangle$ in \eqref{GR2} that is valid for arbitrary $St$, we now develop the result further in order to gain insight into the multiscale nature of the problem. To do this, we introduce the coarse-graining decompositions $u_z(\bm{x},t)=\widetilde{u}_z(\bm{x},t)+u'_z(\bm{x},t)$ and $\bm{\mathcal{V}}=\widetilde{\bm{\mathcal{V}}}+\bm{\mathcal{V}}'$, where $\widetilde{u}_z$ and $\widetilde{\bm{\mathcal{V}}}$ denote the fields $u_z$ and ${\bm{\mathcal{V}}}$ coarse-grained on the length scale $\ell_c(St)$, while $u'_z(\bm{x},t)\equiv u_z(\bm{x},t)-\widetilde{u}_z(\bm{x},t)$ and $\bm{\mathcal{V}}'\equiv\bm{\mathcal{V}}-\widetilde{\bm{\mathcal{V}}}$ are the ``sub-grid'' fields. Inserting these decompositions into \eqref{GR2} we obtain
\begin{align}
\begin{split}
\langle{u}_z(\bm{x}^p(t),t)\rangle=&\Bigg\langle\widetilde{u}_z(\bm{x},t)\exp\Bigg( -\int_0^t {\Big(} 
\bm{\nabla\cdot}\widetilde{\bm{\mathcal{V}}}(\bm{\mathcal{X}}(s\vert\bm{x},t),s)+\bm{\nabla\cdot}\bm{\mathcal{V}}'(\bm{\mathcal{X}}(s\vert\bm{x},t),s) {\Big)} \,ds \Bigg)\Bigg\rangle^{\bm{u}}\\
+&\Bigg\langle{u}'_z(\bm{x},t)\exp\Bigg( -\int_0^t {\Big(}  \bm{\nabla\cdot}\widetilde{\bm{\mathcal{V}}}(\bm{\mathcal{X}}(s\vert\bm{x},t),s)+\bm{\nabla\cdot}\bm{\mathcal{V}}'(\bm{\mathcal{X}}(s\vert\bm{x},t),s) {\Big)} \,ds \Bigg)\Bigg\rangle^{\bm{u}}.
\end{split}
\label{GR2cg}
\end{align} 

We now consider Taylor Reynolds number $R_\lambda\to\infty$, and choose the coarse-graining length scale $\ell_c(St)$ to be a function of $St$, i.e. $\ell_c(St)$. To do this, we define the scale-dependent Stokes number $St_{\ell}\equiv \tau_p/\tau_{\ell}$, where $\tau_{\ell}$ is the eddy-turnover timescale at scale $\ell$. We then define $\ell_c(St)$ through $St_{\ell_c}=\gamma$, where $\gamma$ is a constant such that $\gamma\lll1$. With this definition, $\ell\geq \ell_c(St)$ corresponds to flow scales at which the effects of the particle inertia are negligibly small, and the effects of particle inertia are only felt at scales $\ell < \ell_c(St)$. Since $\tau_{\ell}$ is a non-decreasing function of $\ell$ in homogeneous turbulence, it follows that $\ell_c(St)$ is a non-decreasing function of $St$. In order to illustrate more clearly the connection between $\ell_c$ and $St$, we may use K41 to derive their relationship for the case where $\ell_c$ lies in the inertial range. From K41 we have that for $\ell$ in the inertial range $\tau_\ell\sim\langle\epsilon\rangle^{-1/3}\ell^{2/3}$, and then using the definition of $\ell_c(St)$ we obtain $\ell_c(St)\sim\eta(St/\gamma)^{3/2}$. This then shows how as $St$ is increased, $\ell_c(St)$ also increases.

According to the definition of $\ell_c(St)$, the particle clustering at scales $\ell\geq\ell_c(St)$ is negligible and therefore
\begin{align}
\exp\Bigg( -\int_0^t\bm{\nabla\cdot}\widetilde{\bm{\mathcal{V}}}(\bm{\mathcal{X}}(s\vert\bm{x},t),s) \,ds \Bigg)\approx 1,\label{CGcont}
\end{align}
such that the contribution in \eqref{GR2cg} associated with fluctuations of the particle concentration field $\varphi(\bm{x},t)$ arises only from the sub-grid contribution $\exp( -\int_0^t \bm{\nabla\cdot}\bm{\mathcal{V}}'(\bm{\mathcal{X}}(s\vert\bm{x},t),s) \,ds)$. Further, since $\gamma\lll1$, significant deviations of  $\bm{\nabla\cdot}\bm{\mathcal{V}}'$ from zero will only occur at scales $\ell \ll \ell_c(St)$, and therefore $\bm{\nabla\cdot}\bm{\mathcal{V}}'$ should be uncorrelated with $\widetilde{u}_z(\bm{x},t)$, under the standard assumption that widely separated flow scales in turbulence are uncorrelated. This assumption, together with \eqref{CGcont}, reduces \eqref{GR2cg} to the result
\begin{align}
\begin{split}
\langle{u}_z(\bm{x}^p(t),t)\rangle \approx \Bigg\langle{u}'_z(\bm{x},t)\exp\Bigg( -\int_0^t\bm{\nabla\cdot}\bm{\mathcal{V}}'(\bm{\mathcal{X}}(s\vert\bm{x},t),s)\,ds \Bigg)\Bigg\rangle^{\bm{u}}.
\end{split}
\label{Alt}
\end{align}
Since the RHS of this result only contains the sub-grid fields, it shows that the particle settling speeds are not affected by every scale of the turbulent flow. Instead, only scales of size $\ell <\ell_c(St)$ contribute to the enhanced settling due to turbulence. The physical mechanism embedded in \eqref{Alt} is a multiscale version of the original preferential sweeping mechanism described by \cite{maxey87} and \cite{wang93}. In particular, according to \eqref{Alt}, $\langle{u}_z(\bm{x}^p(t),t)\rangle>0$ occurs because the inertial particles are preferentially swept by eddies of size $\ell <\ell_c(St)$.
\subsection{Implications of result}
A number of interesting and important implications and predictions follow from \eqref{Alt}, which we now discuss.
\subsubsection{The scales of motion that influence the particle settling speed}\label{sec:Theory-Scalesofmotion}
The result in \eqref{Alt} shows that the turbulent flow scales that contribute to the enhanced particle settling speeds are those with size less than $\ell_c(St)$, while scales of size greater than or equal to $\ell_c(St)$ make a negligible contribution. Since $\ell_c(St)$ is a non-decreasing function of $St$ then it follows that increasingly larger scales contribute to the enhanced settling speeds as $St$ is increased. 

One important implication of this is that there cannot, on theoretical grounds, be any single flow scale that determines $\langle{u}_z(\bm{x}^p(t),t)\rangle$. This is in contrast to previous work \citep[e.g.][]{wang93,yang98,good14,nemes17} where it has been argued that the relevant velocity scale determining $\langle{u}_z(\bm{x}^p(t),t)\rangle$ is $u'$, which is associated with the large scales of the flow. We would argue that the results in those previous studies were strongly affected by the fact that the $R_\lambda$ values they considered were such that $u'/u_\eta$ was not very large. 

According to \eqref{Alt}, the flow scales that determine $\langle{u}_z(\bm{x}^p(t),t)\rangle$ depend essentially upon $St$, and when $St\ll 1$, we expect $\ell_c(St)=O(\eta)$ so that $\langle{u}_z(\bm{x}^p(t),t)\rangle$ depends on $u_\eta$, not $u'$. On the other hand, when $1\ll St\ll \tau_L/\tau_\eta$ (where $\tau_L$ is the integral timescale of the flow), the velocity scale dominating $\langle{u}_z(\bm{x}^p(t),t)\rangle$ will correspond to those of inertial range eddies that have velocities much greater than $u_\eta$, but much less than $u'$ (when $R_\lambda\to\infty$).

\subsubsection{Influence of $R_\lambda$ on the particle settling speed}\label{sec:Theory-InfluenceofRe}
\begin{figure}
	\centering
	\vspace{0mm}
	\includegraphics{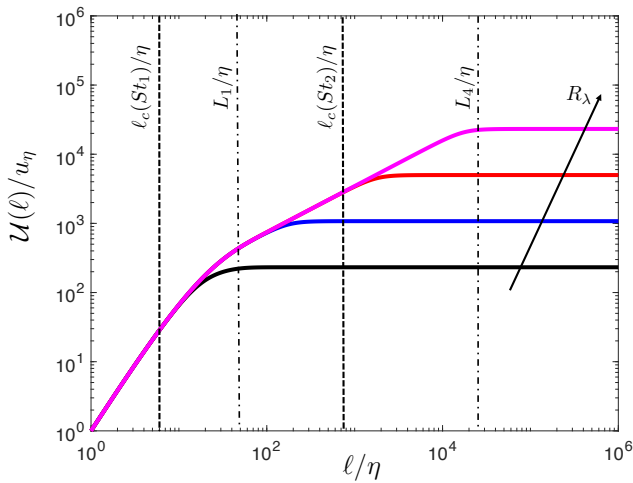}
	\caption{Plot to illustrate how $\ell_c(St)$ determines the $R_\lambda$ dependence of $\langle{u}_z(\bm{x}^p(t),t)\rangle$. The values of $R_\lambda$ shown correspond to $R_\lambda=[29, 133, 615, 2854]$.}
	\label{Fig_01}
\end{figure}

Another implication of \eqref{Alt} concerns the influence of $R_\lambda$ on $\langle{u}_z(\bm{x}^p(t),t)\rangle$. According to K41 \citep{pope}, the velocity scale associated with an eddy of size $\ell$, i.e. $\mathcal{U}(\ell)$, grows as $\mathcal{U}(\ell)\propto \ell$ in the dissipation range, $\mathcal{U}(\ell)\propto \ell^{1/3}$ in the inertial range, and for $\ell>L$ we have $\mathcal{U}(\ell)= u'$, where $L$ is the integral length scale of the flow. Further, using K41 we also have $\mathcal{U}(\ell\geq L)/u_\eta\propto R_\lambda^{1/2}$. In figure \ref{Fig_01}, we plot $\mathcal{U}(\ell)/u_\eta$ for four different values of $R_\lambda$ (these curves are plotted using the curve fit for $\mathcal{U}(\ell)$ in \cite{zaichik09}), and we denote the integral length scales for these flows as $L_1,L_2,L_3,L_4$, where $L_1<L_2<L_3<L_4$ (we assume $\eta$ is the same for each flow for simplicity of the discussion). Also indicated using dashed lines are two values of $\ell_c(St)$ corresponding to Stokes numbers $St_1$ and $St_2$ where $St_1<St_2$ so that $\ell_c(St_1)<\ell_c(St_2)$. An important point is that when normalized by the Kolmogorov scales, $\mathcal{U}(\ell)$ is independent of $R_\lambda$ when $\ell <L_1$. Indeed a consequence of K41 is that when considering any two turbulent flows, $\mathcal{U}(\ell)/u_\eta$ will be independent of $R_\lambda$ up to scales where $\ell$ is of the order of the integral lengthscale of the flow that has the smallest $R_\lambda$.

Now, let us consider how $\langle{u}_z(\bm{x}^p(t),t)\rangle/u_\eta$ would vary across these four $R_\lambda$ values for $St_1$ and $St_2$. For $St_1$, since $\ell_c(St_1)<L_1$ then $\langle{u}_z(\bm{x}^p(t),t)\rangle/u_\eta$ should be independent of $R_\lambda$ since $\mathcal{U}(\ell)/u_\eta$ is the same for each of the flows until $\ell\geq L_1$. On the other hand, for $St_2$, $\langle{u}_z(\bm{x}^p(t),t)\rangle/u_\eta$ will differ for the smallest two values of $R_\lambda$ since $\ell_c(St_2)>L_2$. However, for the largest two values of $R_\lambda$, $\langle{u}_z(\bm{x}^p(t),t)\rangle/u_\eta$ should be the same since  $\ell_c(St_2)<L_3$. In other words, the $R_\lambda$ dependence of  $\langle{u}_z(\bm{x}^p(t),t)\rangle/u_\eta$ for $St_2$ will saturate once the integral lengthscale of the flow, $L(R_\lambda)$, exceeds $\ell_c(St_2)$.

More generally, $\langle{u}_z(\bm{x}^p(t),t)\rangle/u_\eta$ will only depend on $R_\lambda$ while $\ell_c(St)>L(R_\lambda)$. For finite $St$, $\ell_c(St)$ is always finite, and therefore for $R_\lambda\to\infty$, the $R_\lambda$ dependence of $\langle{u}_z(\bm{x}^p(t),t)\rangle/u_\eta$ will always saturate for all finite $St$.

\subsubsection{Influence of Froude number on the scales governing particle settling speeds}

Usually, preferential concentration at scale $\ell$ is said to be strongest when $St_\ell=O(1)$. However, in the presence of gravity, this needs to be nuanced. In particular, we should instead say that preferential concentration at scale $\ell$ is strongest when $St\tau_\eta=O(T_\ell)$, where $T_\ell$ is the eddy turnover timescale \emph{seen by the particle}, which depends upon $St$ and $Fr$, and, like $\tau_\ell$, is a non-decreasing function of $\ell$. For a given $St$ and $\ell$, $T_\ell$ decreases with decreasing $Fr$, reflecting the fact that the faster a particle falls through the flow, the faster the fluid velocity along its trajectory will decorrelate. This then means that for a given $St$, as $Fr$ is decreased, one has to go to larger flow scales in order to observe $St\tau_\eta=O(T_\ell)$. Now suppose we re-define $\ell_c(St)$ using $T_\ell$ instead of $\tau_\ell$, by defining $\ell_c(St)$ through $St\tau_\eta/T_{\ell_c}=\gamma$, and again take $\gamma\lll1$. With this definition we observe that for fixed $St$, $\ell_c(St)$ increases with decreasing $Fr$, i.e. as $Fr$ is decreased, one has to go to larger values of $\ell$ in order to satisfy $St\tau_\eta/T_{\ell}=\gamma$. Through \eqref{Alt}, this then means that as $Fr$ is decreased, the scales contributing to the particle settling enhancement increase. 

\section{Testing the arguments using DNS data}\label{sec:DNS_Details}

We will test the theoretical arguments presented in the previous section using DNS data spanning a range of $R_\lambda$, $St$ and $Fr$.

\subsection{DNS Details}

Our DNS dataset is identical to that in \citet{ireland16a,ireland16b}, and we therefore refer the reader to that paper for the details of the DNS. Here we just give a brief summary. We perform a pseudo-spectral DNS of isotropic turbulence on a three-dimensional triperiodic cube of length $\mathcal{L}$ with $N^3$ grid points. The Navier-Stokes equation for an incompressible fluid was solved, with the pressure term eliminated using the divergence-free condition for the velocity field.
\begingroup
\begin{align}
\bm{\nabla \cdot {u} } = 0 
\end{align}
\vspace{-5mm}
\begin{align}
\frac{\partial {\bm u}}{\partial t} + \bm {\omega \times \bm u} + \bm{\nabla} \Bigg( \frac{P}{\rho_f} + \frac{u^2}{2} \Bigg) = \nu \nabla^{2} \bm u + \bm f
\end{align}
\endgroup
where $\bm u$ is the fluid velocity, $\bm {\omega}$ is the voticity, $P$ is the pressure, $\rho_f$ is the fluid density, $\nu$ is the kinematic viscosity and $\bm f$ is a large-scale forcing term that is added to make the flow field statistically stationary.  Deterministic forcing was applied to wavenumbers with magnitude $k = \sqrt{2}$. A detailed description of the numerical methods used can be found in \citet{ireland13}. The gravity term in the Navier-Stokes equation is precisely cancelled by the mean pressure gradient and so it has no dynamical consequence on the turbulent flow field. When using periodic boundary conditions, particles can
artificially re-encounter the same large eddy as they are periodically looped through the domain, if the time it takes the settling particles to traverse the distance $\mathcal{L}$ is smaller than the large eddy turnover time, i.e if $\mathcal{L}/St\tau_\eta g < O(\tau_L)$. In all of our simulations, the domain lengths $\mathcal{L}$ are chosen to satisfy $\mathcal{L}/St\tau_\eta g >\tau_L$, thereby suppressing the artificial periodicity effects for settling particles (a systematic study of this was presented in \cite{ireland16b}).

In order to analyze the same dynamical system we considered in the theory, we track inertial particles governed by (\ref{eom}), where particle are acted upon by both gravity and linear drag force.  We assume that the particle loading is dilute and hence inter particle interactions and two-way coupling can be neglected \citep{elghobashi93,sundaram97}. An eight-point B-spline interpolation scheme (with $C^6$ continuity) based on the algorithm in \citet{vanhinsberg12} was used to compute the fluid velocity at the particle position, ${u}(\bm{x}^p(t),t)$. We consider $R_\lambda=90,230$ and $398$ (note that we refer to the simulations with $R_\lambda=224,230,237$ nominally as having $R_\lambda\approx 230$ since they differ only due to statistical fluctuations, but correspond to the same $\nu$) and two $Fr$ that are representative of the conditions in clouds, $Fr = 0.052$ for a cumulus clouds and $Fr = 0.3$ for a strongly turbulent cumulonimbus cloud \citep[see][]{ireland16b}. Fourteen different particle classes are simulated with $St\in[0,2]$. Details of the simulations are given in table \ref{Table01}. Note that the data for $R_\lambda =237$ and $Fr=0.3$ is from \citet{momenifar18}, while the rest is from \citet{ireland16a,ireland16b}.

\begin{table}
\begin{center}
\begin{tabular}{c c c c c c c}
Simulation              & I             & II            & III           & IV           & V           & VI            \\
$R_{\lambda}$           & 90            & 224           & 230           & 237          & 398         & 398           \\
$Fr$                    & 0.052         & $\infty$      & 0.052         & 0.3          & $\infty$    & 0.052         \\
$\mathcal{L}$           & $16\pi$       & $2\pi$        & $4\pi$        & $2\pi$       & $2\pi$      & $2\pi$        \\
$N$                     & 1024          & 512           & 1024          & 512          & 1024        & 1024          \\
$\nu$                   & 0.005         & 0.0008289     & 0.0008289     & 0.0008289    & 0.0003      & 0.0003        \\
$\langle\epsilon\rangle$              & 0.257         & 0.253         & 0.239         & 0.2842       & 0.223       & 0.223         \\
$L$                     & 1.47          & 1.40          & 1.49          & 1.43         & 1.45        & 1.45          \\
$L/\eta$                & 55.6          & 204           & 213           & 214          & 436         & 436           \\
$u'$                    & 0.912         & 0.915         & 0.914         & 0.966        & 0.915       & 0.915         \\
$u'/u_{\eta}$           & 4.82          & 7.60          & 7.70          & 7.82         & 10.1        & 10.1          \\
$\tau_L$                & 1.61          & 1.53          & 1.63          & 1.48         & 1.58        & 1.58          \\
$\tau_L/\tau_{\eta}$    & 11.52         & 26.8          & 27.66         & 27.36        & 43.0        & 43.0          \\
$k_{max}\eta$           & 1.61          & 1.66          & 1.68          & 1.62         & 1.60        & 1.60          \\
$N_p$                   & 16,777,216    & 2,097,152     & 16,777,216    & 2,097,152    & 2,097,152   & 2,097,152     \\

\end{tabular}

\caption{Flow parameters in DNS of isotropic turbulence where all dimensional parameters are in arbitrary units and all statistics are averaged over the duration of the run ($T$). $R_{\lambda} \equiv u'\lambda/\nu \equiv 2k/\sqrt{5/3\nu \langle\epsilon\rangle} $ is the Taylor microscale Reynolds Number,  $Fr$ is the Froude number, $\lambda$ is the Taylor microscale, $\mathcal{L}$ is the domain size, $N$ is the number of grid points in each direction, $\nu$ is the fluid kinematic viscosity, $\langle\epsilon\rangle \equiv 2 \nu \int_{0}^{\kappa_{max}} \kappa^2 E(\kappa) d\kappa$ is the mean turbulent kinetic energy dissipation rate, $\kappa$ is the wavenumber in Fourier space, $E$ is the energy spectrum, $L \equiv (3\pi/2k) (\int_{0}^{\kappa_{max}} (E(\kappa)/\kappa) d\kappa)  $ is the integral length scale, $\eta \equiv (\nu^{3}/\langle\epsilon\rangle)^{1/4} $ is the Kolmogorov length scale, $u' \equiv \sqrt{2k/3}$ is the r.m.s of fluctuating fluid velocity, $k$ is the turbulent kinetic energy, $u_{\eta} \equiv (\langle\epsilon\rangle \nu)^{3/4} $ is the Kolmogorov velocity scale, $\tau_L \equiv L/u'$ is the large-eddy turnover time, $\tau_{\eta} \equiv \sqrt{\nu/\langle\epsilon\rangle}$ is the Kolmogorov time scale, $k_{max} = \sqrt{2N/3}$ is the maximum resolved wavenumber and $N_p$ is the number of particles per Stokes number used in the simulation. The grid spacing is kept constant as the domain size is increased, and so the small-scale resolution, $k_{max}\eta$, is approximately constant between the different domain sizes. The viscosity and forcing parameters were kept the same when increasing domain size, and thus both small-scale and large-scale flow parameters are held approximately constant. }
\label{Table01}
\end{center}
\end{table}

\subsection{Testing Methodology}

It would be very difficult to directly test \eqref{Alt} owing to the practical difficulty in constructing the field $\bm{\mathcal{V}}'(\bm{x},t)$ (and a simple evolution equation for $\bm{\mathcal{V}}'(\bm{x},t)$ is not available). However, the insights and predictions from the theoretical analysis can be tested by computing $\langle{{u}'_z}(\bm{x}^p(t),t)\rangle$ for various coarse-graining length scales, and analyzing how the coarse-graining affects the results for varying $St, R_\lambda$ and $Fr$. Strictly speaking, $\langle{{u}'_z}(\bm{x}^p(t),t)\rangle$ does not actually correspond to \eqref{Alt}, but rather corresponds to \eqref{Alt} with $\bm{\nabla\cdot}\bm{\mathcal{V}}'$ replaced by $\bm{\nabla\cdot}\bm{\mathcal{V}}$. However, according to the arguments and definitions leading to \eqref{Alt}, these two quantities are asymptotically equivalent since the coarse-graining length scale $\ell_c$ is \emph{defined} in such a way that $\bm{\nabla\cdot}\bm{\mathcal{V}}'\approx \bm{\nabla\cdot}\bm{\mathcal{V}}$ since $\vert\bm{\nabla\cdot}\widetilde{\bm{\mathcal{V}}}\vert\lll \vert\bm{\nabla\cdot}\bm{\mathcal{V}}'\vert$. Therefore, comparing $\langle{{u}'_z}(\bm{x}^p(t),t)\rangle$ with the implications and predictions following from \eqref{Alt} is appropriate.

To take full advantage of our existing large-database on inertial particle motion in isotropic turbulence \citep[see][]{ireland16a,ireland16b}, we compute $\langle{{u}'_z}(\bm{x}^p(t),t)\rangle$ from our existing DNS data via postprocessing. To do this we take our DNS data for ${u}_z(\bm{x},t)$ and the particle positions $\bm{x}^p(t)$ at a number of different times, for multiple $St$, $R_\lambda$ and $Fr$. We then apply a sharp spectral cut-off at wavenumber $k_F$ to ${u}_z(\bm{x},t)$, and from this obtain the sub-grid field through
\begin{align}
{u}'_z(\bm{x},t)\equiv u_z(\bm{x},t)-\widetilde{u}_z(\bm{x},t)=\Large\sum_{\|\bm{k}\|> k_F}\hat{u}_z(\bm{k},t)e^{i\bm{k\cdot x}},
\end{align}
where here and throughout, $\widetilde{(\cdot)}$ denotes the coarse-grained field, while $(\cdot)'$ denotes the sub-grid field. We then interpolate $u'_z(\bm{x},t)$ to the positions of inertial particles $\bm{x}^p(t)$ using an eight-order B-spline interpolation scheme to obtain ${u}'_z(\bm{x}^p(t),t)$. The values of ${u}'_z(\bm{x}^p(t),t)$ are then averaged over all the particles (with a given $St$) and over multiple times to obtain $\langle{{u}'_z}(\bm{x}^p(t),t)\rangle$. This process is then repeated for multiple $k_c$ in order to examine the effect of the coarse-graining and how its effect depends on $St,R_\lambda$ and $Fr$. In order to relate the spectral cut-off wavenumber $k_F$ to a physical space filtering scale we define $\ell_F\equiv 2\pi/k_F$ \citep{eyink09}. 

By considering the results of $\langle{{u}'_z}(\bm{x}^p(t),t)\rangle$ for various $St,\ell_F,Fr$ and $R_\lambda$ and comparing them with those for $\langle{{u}_z}(\bm{x}^p(t),t)\rangle$, we can test the predictions of the theory regarding which flow scales contribute to $\langle{{u}_z}(\bm{x}^p(t),t)\rangle$.
\section{Results and discussion}\label{sec:DNS_Results}
\subsection{The scales of motion that influence the particle settling speed}
\begin{figure}
	\centering
	\vspace{0mm}
    \includegraphics{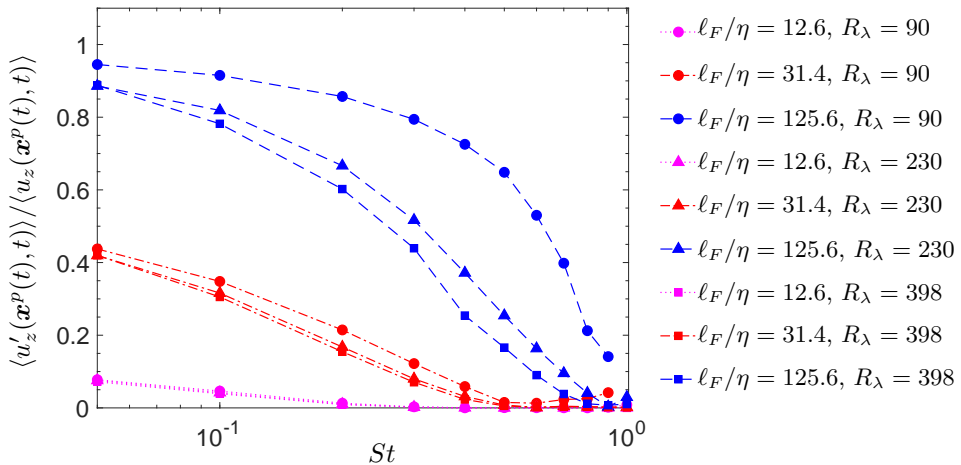}
	\caption{DNS results for $\langle u'_z(\bm{x}^p(t),t)\rangle/\langle u_z(\bm{x}^p(t),t)\rangle$ as a function of $St$, for $Fr=0.052$, and for various filtering lengths $\ell_F/\eta$. The circles correspond to $R_\lambda=90$, the triangles to $R_\lambda\approx 230$, and the squares to $R_\lambda=398$. The dashed lines correspond to $\ell_F/\eta=125.6$, the dash-dot lines to $\ell_F/\eta=31.4$, and the dotted lines to $\ell_F/\eta=12.6$.}
	\label{Fig_02}
\end{figure}

The theoretical analysis predicts that the range of scales that contribute to $\langle{{u}}_z(\bm{x}^p(t),t)\rangle$ should monotonically increase as $St$ increases. To test this prediction, in figure \ref{Fig_02}, we plot the ratio $\langle u'_z(\bm{x}^p(t),t)\rangle/\langle{{u}}_z(\bm{x}^p(t),t)\rangle$ for various filtering length scales $\ell_F$, and various $R_\lambda$. For $St\lll1$ we would have $\ell_c(St)=O(\eta)$, and so filtering out scales $\ell_F>O(\eta)$ would have little effect, since the particle settling speed is only affected by the scales less than $\ell_c(St)$. Hence, for $St\lll1$, we expect $\langle u'_z(\bm{x}^p(t),t)\rangle/\langle{{u}}_z(\bm{x}^p(t),t)\rangle \approx 1$. On the other hand, for $St=O(1)$ we would have $\ell_c(St)>O(\eta)$, and so filtering out scales $\ell_F>O(\eta)$ would have a strong effect, since these scales make a strong contribution to the particle settling speed. Hence, for $St=O(1)$, we expect $\langle u'_z(\bm{x}^p(t),t)\rangle/\langle{{u}}_z(\bm{x}^p(t),t)\rangle \ll 1$. More generally, the prediction is that for a given $\ell_F$, $\langle u'_z(\bm{x}^p(t),t)\rangle/\langle{{u}}_z(\bm{x}^p(t),t)\rangle$ should decrease with increasing $St$, reflecting the fact that $\langle{{u}}_z(\bm{x}^p(t),t)\rangle$ is affected by increasingly larger scales as $St$ is increased. The results in figure \ref{Fig_02} confirm this prediction. They also reveal how sensitive $\langle{{u}}_z(\bm{x}^p(t),t)\rangle$ is to scales $\ell \gg \eta$, even when $St=O(0.1)$, which is quite surprising.

\begin{figure}
	\centering
	\vspace{0mm}
    \includegraphics{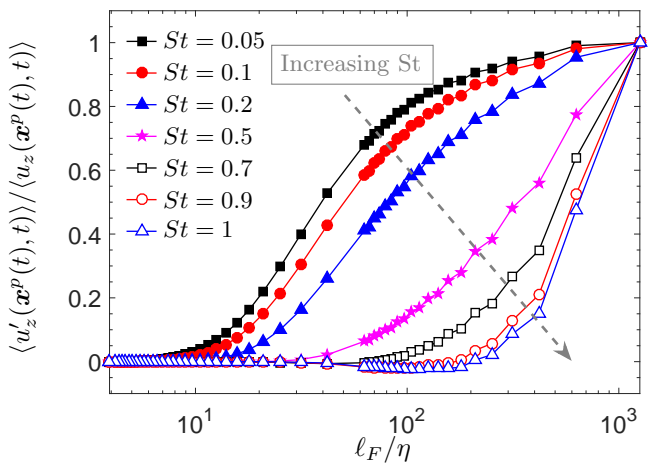}
	\caption{DNS data for $\langle u'_z(\bm{x}^p(t),t)\rangle/\langle u_z(\bm{x}^p(t),t)\rangle$ as a function of $\ell_F/\eta$ for different $St$, and $Fr=0.052$, $R_\lambda=398$.}
	\label{Fig_03}
\end{figure}

To further illustrate this behavior, in figure \ref{Fig_03}, we plot $\langle u'_z(\bm{x}^p(t),t)\rangle/\langle u_z(\bm{x}^p(t),t)\rangle$ as a function of $\ell_F/\eta$ for various $St$. For $St\lesssim 0.2$ we clearly see that as $\ell_F$ is increased, $\langle u'_z(\bm{x}^p(t),t)\rangle$ approaches a constant value, implying that there do indeed exist scales beyond a certain size that have a negligible effect on the particle settling velocity, as predicted by the theory. However, for $St=O(1)$, we do not see such an asymptote, and their settling velocity is significantly affected by the largest scales in the flow. In order to understand why this is the case, we will now estimate $\ell_c(St)$. 

\begin{figure}
	\centering
	\vspace{0mm}
    \includegraphics{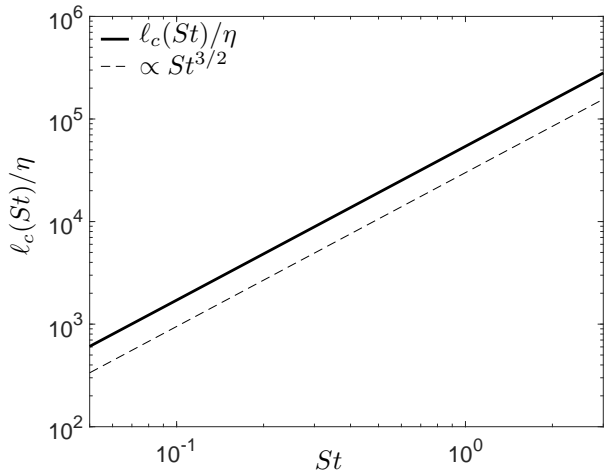}
	\caption{Plot illustrating how $\ell_c(St)/\eta$ grows with $St$.}
	\label{lc_plot}
\end{figure}
%

Recall that in the derivation of our theoretical result we prescribed the parameter $\gamma$ to have the asymptotic value $\gamma \lll 1$. However, we may estimate a value for $\gamma$ from our DNS. To do this, we note that for $St=0.05$, $\langle u'_z(\bm{x}^p(t),t)\rangle/\langle u_z(\bm{x}^p(t),t)\rangle\approx 1$ for $\ell_F/\eta\gtrsim 600$ and hence $\ell_c(St=0.05)\approx 600\eta$. Then using the K41 result discussed earlier, $\ell_c(St)\sim\eta(St/\gamma)^{3/2}$ and using $\ell_c(St=0.05)\approx 600\eta$, we obtain the estimate $\gamma\approx 7\times 10^{-4}$. Using this value, in figure~\ref{lc_plot} we plot $\ell_c(St)\sim\eta(St/\gamma)^{3/2}$, where we have taken $R_\lambda\to\infty$ so that the inertial range scaling $\ell_c(St)\sim\eta(St/\gamma)^{3/2}$ applies for all $St\geq 0.05$. Using this estimated behavior we find that for $St=1$, $\ell_c(St)/\eta\approx 5.4\times 10^{4}$. In our DNS at $R_\lambda=398$, the ratio of the integral length scale $L$ to $\eta$ is $L/\eta=4.3\times 10^{2}$ which is two orders of magnitude smaller than $\ell_c(St)/\eta$ for $St=1$. This then explains why in our DNS we do not observe a saturation of $\langle u'_z(\bm{x}^p(t),t)\rangle/\langle u_z(\bm{x}^p(t),t)\rangle$ for $St=O(1)$ as $\ell_F$ is increased. An important conclusion that follows from this is that while the settling speed of particles is dominated by a restricted range of scales, namely scales of size less than $\ell_c(St)$, this range can actually be quite large. Indeed, figure~\ref{lc_plot}, and the results in figure~\ref{Fig_03} indicate that even for $St=0.1$, $\ell_c(St)/\eta\approx 1.7\times 10^{3}$ such that their settling speeds are affected by scales much larger than those in the dissipation range.

These findings also explain why in many previous numerical and experimental studies, it was found that $\langle u_z(\bm{x}^p(t),t)\rangle$ had a strong dependence on $u'$, the large scale fluid velocity scale. We have argued that on theoretical grounds, $\langle u_z(\bm{x}^p(t),t)\rangle$ cannot be characterized by a single flow scale since the range of scales contributing to $\langle u_z(\bm{x}^p(t),t)\rangle$ depend on $St$. However, in these previous works $R_\lambda$ was sufficiently small so that the particle settling speeds were significantly affected by all scales, and as a result $\langle u_z(\bm{x}^p(t),t)\rangle$ was found to have a strong dependence on $u'$. In contrast, in natural flows where $R_\lambda$ is much larger, this would not be the case. In the atmosphere, typical values are $\eta=O(mm)$ and $L=O(100m)$ \citep{shaw03,grabowski13}, and together with the results in figure~\ref{lc_plot} this implies we would have, for example, $\ell_c(St=0.1)\approx 1.7m$ and $\ell_c(St=1)\approx 54m$. Consequently, for $St\leq1$, the large scale fluid velocities in the atmosphere, characterized by $u'$, would play no role in the particle settling. Nevertheless, the estimate $\ell_c(St=1)\approx 54m$ shows that the range of atmospheric flow scales that may contribute to the enhanced settling speeds due to turbulence is quite large. This means that for $St=O(1)$, particle settling in the atmosphere may be strongly influenced by non-ideal effects such as flow inhomogeneity, anisotropy, and stratification (noting that the Ozmidov scale is greater than or equal to $O(m)$ in the atmosphere \citep{riley12}).

So far we have emphasized that as $St$ is increased, larger scales contribute to the particle settling since $\ell_c(St)$ is a non-decreasing function of $St$. However, in reality, as $St$ is increased, not only do larger scales contribute to the particle settling, but smaller scales begin to contribute less. This is because the preferential sweeping effect at any scale is only effective when $St_\ell \not\lll 1$ and $St_\ell\not\ggg 1$. Let us define $\widehat{\ell}_c(St)$ as the scale below which the preferential sweeping mechanism is not effective, so that scales $\ell<\widehat{\ell}_c(St)$ correspond to scales at which $St_\ell\ggg 1$. Then, the scales at which the preferential sweeping mechanism would operate are $\widehat{\ell}_c(St)\leq \ell<\ell_c(St)$, and both $\widehat{\ell}_c(St)$ and ${\ell}_c(St)$ are non-decreasing functions of $St$. While our theoretical analysis could be extended to also include the lower limit scale $\widehat{\ell}_c(St)$, we have chosen not to do so since it would render the theoretical result much more complicated. Furthermore, our principle concern in this paper is to understand the upper limit of the turbulent flow scales that contribute to the particle settling velocity in a turbulent flow, and this is given by ${\ell}_c(St)$. Nevertheless, understanding the minimum flow scales affecting the settling process is key to the development of particle SubGrid Scale (SGS) models for LES of particles settling in turbulent flows. The effect of the smallest scales of the turbulence on particle settling speeds was investigated by \citet{rosa17} using DNS and their results showed that scales smaller than a certain size did not affect the particle settling speeds. Although their data does not provide enough information to determine exactly how this ``cut-off scale'' depends on $St$, their data is consistent with our theoretical prediction that $\widehat{\ell}_c(St)$ is a non-decreasing function of $St$.

\subsection{Influence of $R_\lambda$ on the particle settling speed}

\begin{figure}
	\centering
	\vspace{0mm}
    \includegraphics{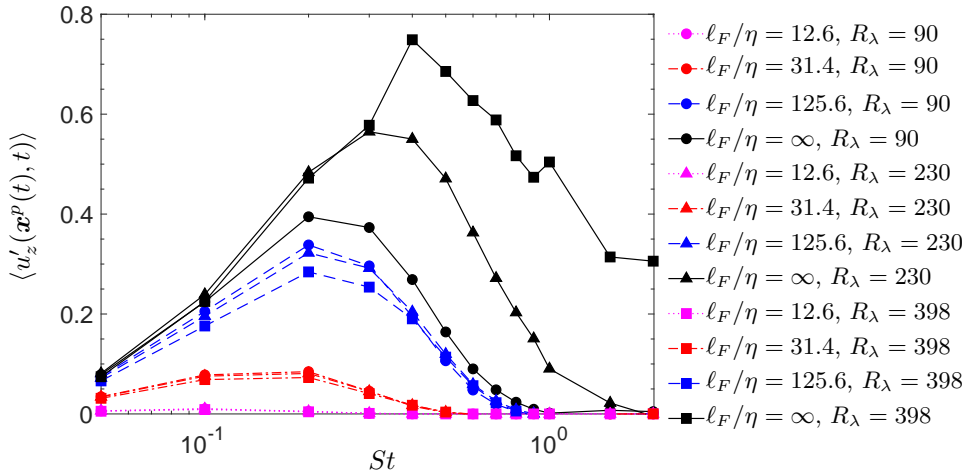}
	\caption{DNS results for $\langle u'_z(\bm{x}^p(t),t)\rangle$ as a function of $St$, for $Fr=0.052$, and for various filtering lengths $\ell_F/\eta$. The circles correspond to $R_\lambda=90$, the triangles to $R_\lambda\approx 230$, and the squares to $R_\lambda=398$. The solid lines correspond to $\ell_F/\eta=\infty$, the dashed lines to $\ell_F/\eta=125.6$, the dash-dot lines to $\ell_F/\eta=31.4$, and the dotted lines to $\ell_F/\eta=12.6$.}
	\label{Fig_04}
\end{figure}

As discussed in \cref{sec:Theory-InfluenceofRe}, as $R_\lambda$ is increased, the range of scales in the turbulent flow increases. According to our theoretical analysis, when $St\lll 1$, $\ell_c(St)$ is small enough so that the particles are not influenced by the additional scales introduced by increasing $R_\lambda$. Consequently, $\langle u_z(\bm{x}^p(t),t)\rangle$ should not vary with $R_\lambda$ for $St\lll 1$. However, as $St$ is increased, so also does $\ell_c(St)$, and for sufficiently large $St$, this allows the particles to feel the effects of the additional flow scales introduced by increasing $R_\lambda$. As a result, $\langle u'_z(\bm{x}^p(t),t)\rangle$ can depend on $R_\lambda$ as $St$ is increased. The results in figure \ref{Fig_04} confirm this picture and show that without filtering (i.e. $\ell_F/\eta=\infty$), $\langle u'_z(\bm{x}^p(t),t)\rangle=\langle u_z(\bm{x}^p(t),t)\rangle$ is significantly enhanced with increasing $R_\lambda$ when $St\gtrsim 0.2$, whereas it is almost insensitive to $R_\lambda$ when $St\lesssim 0.1$.

The results in figure \ref{Fig_04} also show that when scales larger than some finite $\ell_F/\eta$ are filtered out, so that the range of scales contained in $u'_z$ is the same for each $R_\lambda$, the $R_\lambda$  dependence of $\langle u'_z(\bm{x}^p(t),t)\rangle$ is dramatically suppressed. Indeed, for $St\gtrsim 0.4$ the effect of $R_\lambda$ is entirely suppressed for $\ell_F/\eta$ values considered. This confirms that the strong effect of $R_\lambda$ on $\langle{{u}}_z(\bm{x}^p(t),t)\rangle$ is principally due to the enhanced range of scales available for the particles to preferentially sample as $R_\lambda$ is increased. Recall that increasing $R_{\lambda}$ leads to two distinct effects, namely an increased range of flow scales, and enhanced intermittency. While filtering eliminates the effect of the increased range of scales by removing scales greater than $\ell_F$ (so that each flow of differing $R_{\lambda}$ has the same range of scales), the effect of enhanced intermittency still remains. It can be seen in figure \ref{Fig_04} that for $\ell_F/\eta \leq 125.6$, the curves collapse for $St\gtrsim 0.4$, while there is a residual effect of $R_\lambda$ for $St\lesssim 0.3$, which must be due to intermittency. That the effect of intermittency is only apparent for small $St$ is consistent with previous works which show that the effect of flow intermittency on inertial particle motion in turbulence is mainly confined to small $St$, while larger $St$ particles filter out the effects of intermittent fluctuations in the flow \citep{bec06a,sathya08a}.

\begin{figure}
	\centering
	\vspace{0mm}
    \includegraphics{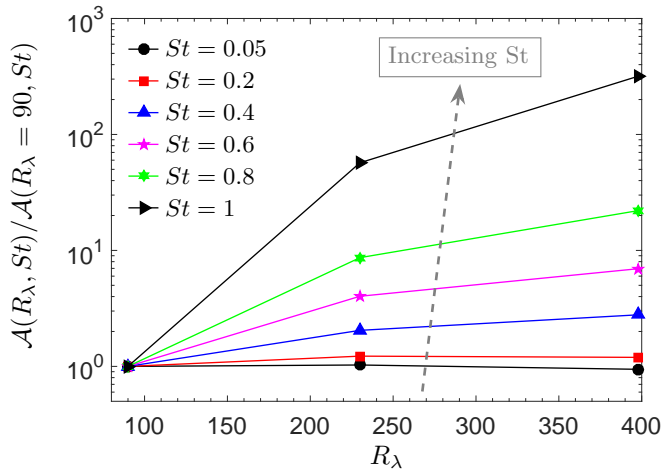}
	\caption{DNS results for $\mathcal{A}(R_\lambda, St)/\mathcal{A}(R_\lambda=90, St)$ as a function of $R_\lambda$, for various $St$, and for $Fr=0.052$.}
	\label{Fig_05}
\end{figure}

To observe the effect of $R_\lambda$ more clearly, in figure \ref{Fig_05}, we plot $\mathcal{A}(R_\lambda, St)/\mathcal{A}(R_\lambda=90, St)$ where $\mathcal{A}(R_\lambda, St)\equiv\langle u_z(\bm{x}^p(t),t)\rangle/u_\eta$, as a function of $R_\lambda$ and for different values of $St$. In agreement with the theoretical analysis, as $St$ is increased, $\mathcal{A}(R_\lambda, St)/\mathcal{A}(R_\lambda=90, St)$ becomes increasingly sensitive to $R_\lambda$. The analysis leads us to expect that for any $St$, the ratio $\mathcal{A}(R_\lambda, St)/\mathcal{A}(R_\lambda=90, St)$ will eventually saturate at sufficiently large $R_\lambda$ and the $R_\lambda$ at which saturation occurs would increase with $St$. Our data is consistent with this, however, we do not have enough $R_\lambda$ data points to be conclusive, and data at larger $R_\lambda$ is required to observe the saturation for $St=O(1)$. Again, this is because in order to observe the saturation we must consider values of $R_\lambda$ for which $\ell_c(St)<L(R_\lambda)$, and our DNS does not satisfy this for $R_\lambda\leq 398$ and $St=O(1)$.

\subsection{Influence of Froude number on the scales governing particle settling speeds}%

\begin{figure}
	\centering
	\vspace{0mm}
    \includegraphics{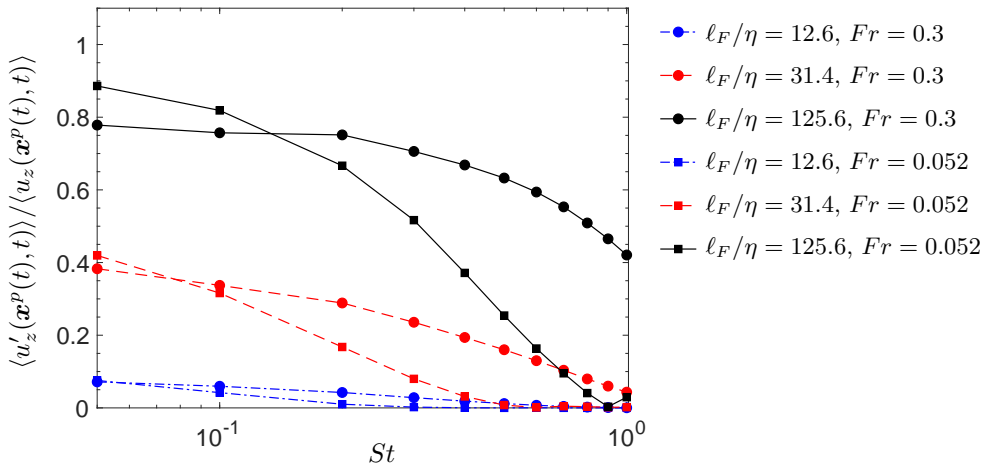}
	\caption{DNS results for $\langle u'_z(\bm{x}^p(t),t)\rangle/\langle{{u}}_z(\bm{x}^p(t),t)\rangle$ as a function of $St$ for $Fr=0.052$ (squares) and $Fr=0.3$ (circles), and $R_\lambda\approx 230$. The solid lines correspond to $\ell_F/\eta=125.6$, the dashed lines to $\ell_F/\eta=31.4$, and the dash-dot lines to $\ell_F/\eta=12.6$.}
	\label{Fig_06}
\end{figure}

Another prediction of the theory is that for a given $St$, as $Fr$ is decreased, $\ell_c(St)$ increases meaning that larger scales become responsible for the behavior of $\langle{{u}}_z(\bm{x}^p(t),t)\rangle$. In figure \ref{Fig_06}, we plot $\langle u'_z(\bm{x}^p(t),t)\rangle$ for $Fr=0.3$ and $Fr=0.052$, each at $R_\lambda\approx 230$. For $St\gtrsim 0.1$, the results confirm the prediction, since they show that $\langle{{u}}_z(\bm{x}^p(t),t)\rangle$ is more strongly affected by filtering as $Fr$ is decreased, which is equivalent to saying that $\langle{{u}}_z(\bm{x}^p(t),t)\rangle$ is affected by increasingly larger scales as $Fr$ is decreased. 

\begin{figure}
	\centering
	\vspace{0mm}
    \includegraphics{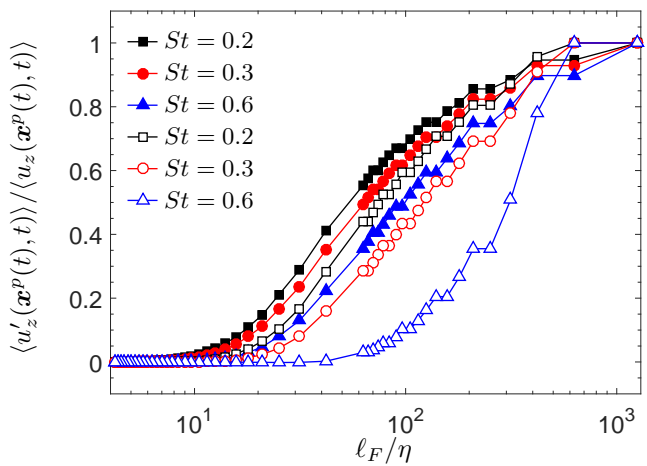}
	\caption{DNS data for $\langle u'_z(\bm{x}^p(t),t)\rangle/\langle u_z(\bm{x}^p(t),t)\rangle$ as a function of $\ell_F/\eta$ for different $St$, and $Fr=0.3$ (filled symbols), $Fr=0.052$ (open symbols), and $R_\lambda\approx 230$.}
	\label{uratio_vs_ellf_Re230}
\end{figure}

To show this more clearly, in figure~\ref{uratio_vs_ellf_Re230} we plot the ratio $\langle u'_z(\bm{x}^p(t),t)\rangle/\langle u_z(\bm{x}^p(t),t)\rangle$ as a function of $\ell_F/\eta$, for different $Fr$, $St$ and for $R_\lambda\approx 230$. The results show that as $Fr$ is decreased, the ratio decreases for a given $St$ and $\ell_F/\eta$. This confirms the prediction of the theory as it indicates that as $Fr$ is decreased, larger scales become responsible for the enhanced particle settling speeds due to turbulence. This could also explain the increased cluster size (determined by Voronoi tessellation) in the presence of gravity observed by \citet{baker17}. In particular, as the particles settle faster due to gravity, the scales at which the clustering mechanisms (such as the preferential sampling of the filtered velocity gradient field, see \cite{bragg14e}) become active move to larger scales.

\subsection{Scale dependence of preferential sweeping}\label{pref_samp}
So far, we have tested the predictions following from our theoretical analysis regarding the scales that contribute to enhanced particles settling speeds in turbulence. We now turn to consider in more detail the multiscale nature of the preferential sweeping mechanism. Recall that the preferential sweeping mechanism involves the idea that the enhanced particle settling is associated with the tendency of inertial particles to preferentially sample the flow by preferring paths around the downward side of vortices. However, according to our theoretical analysis this preferential sweeping can only take place at scales $\ell < \ell_c(St)$, such that the preferential sweeping mechanism is multiscale in general, and does not only involve the small scales as in the $St\ll 1$ analysis of \cite{maxey87}.

A traditional way to consider preferential sampling of the flow by inertial particles is to consider $\mathcal{Q}^p(t)\equiv \mathcal{S}^2(\bm{x}^p(t),t)-\mathcal{R}^2(\bm{x}^p(t),t)$, where $\mathcal{S}^2$ and $\mathcal{R}^2$ are the second invariants of the strain-rate $\bm{\mathcal{S}}\equiv(\bm{\nabla u}+\bm{\nabla u}^\top)/2$ and rotation-rate $\bm{\mathcal{R}}\equiv(\bm{\nabla u}-\bm{\nabla u}^\top)/2$ tensors, respectively. Comparing the statistics of $\mathcal{Q}^p(t)$ along fluid and inertial particle trajectories provides a clear way to consider preferential sampling since their statistics can only differ if the inertial particles are both non-uniformly distributed, and if their distribution is correlated to the local flow, i.e. if they exhibit preferential concentration. However, in order to consider how the particles preferential sample the flow at different scales we must instead consider the coarse-grained quantity $\widetilde{\mathcal{Q}}^p(t)\equiv \widetilde{\mathcal{S}}^2(\bm{x}^p(t),t)-\widetilde{\mathcal{R}}^2(\bm{x}^p(t),t)$, where $\widetilde{\mathcal{S}}^2\equiv \widetilde{\bm{\mathcal{S}}}\bm{:}\widetilde{\bm{\mathcal{S}}}$, $\widetilde{\mathcal{R}}^2\equiv \widetilde{\bm{\mathcal{R}}}\bm{:}\widetilde{\bm{\mathcal{R}}}$, and $\widetilde{\bm{\mathcal{S}}}$, $\widetilde{\bm{\mathcal{R}}}$ denote $\bm{\mathcal{S}}$, $\bm{\mathcal{R}}$ coarse-grained on the scale $\ell_F$.  

\begin{figure}
	\centering
	\vspace{0mm}
	\subfloat[]
	{\includegraphics{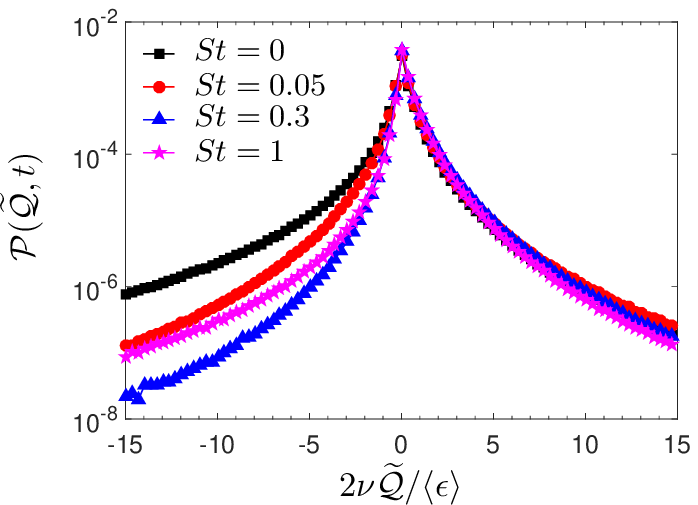} \vspace{8mm}}
		\subfloat[]
	{\includegraphics{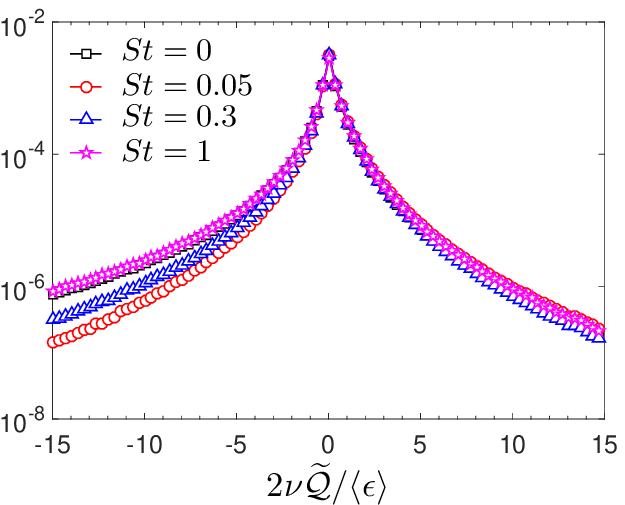} }\\
		\subfloat[]
	{\includegraphics{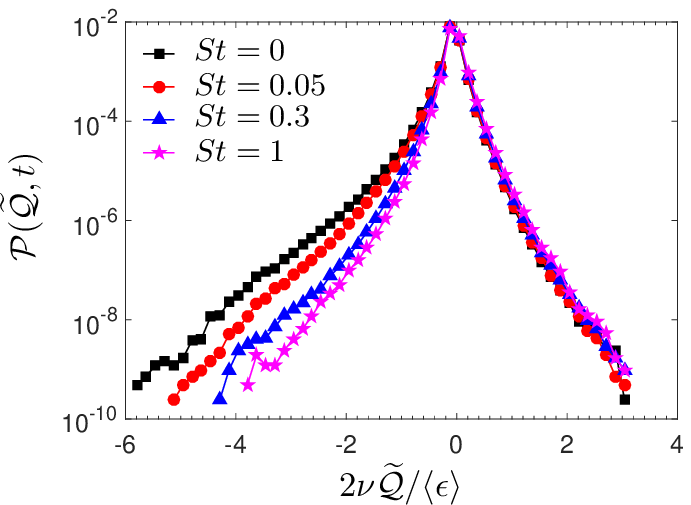} \vspace{8mm}}
		\subfloat[]
	{\includegraphics{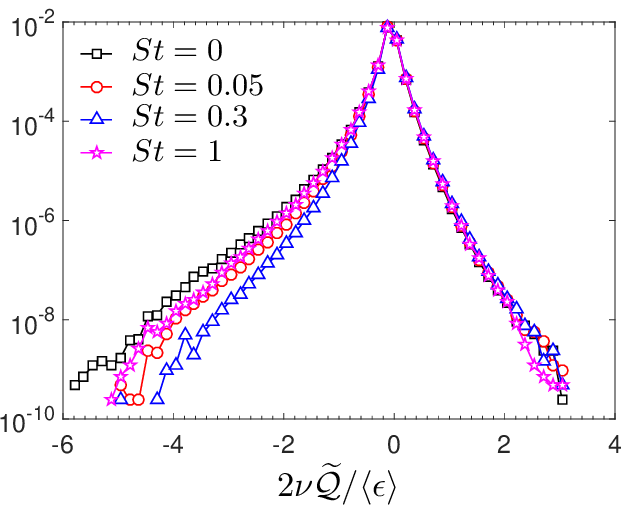}}
	\caption{DNS data for $\mathcal{P}(\widetilde{\mathcal{Q}},t)$ and different $St$. Plots (a),(b) are for $\ell_F/\eta=0$, while plots (c),(d) are for $\ell_F/\eta=66.1$. Plots (a),(c) are for $Fr=\infty$, while plots (b),(d) are for $Fr=0.052$.}
	\label{PDF_Q}
\end{figure}

In figure \ref{PDF_Q}, we plot the PDF of $\widetilde{\mathcal{Q}}^p(t)$, namely
\begin{align}
    \mathcal{P}(\widetilde{\mathcal{Q}},t)\equiv\langle\delta(\widetilde{\mathcal{Q}}^p(t)-\widetilde{\mathcal{Q}})\rangle,
\end{align}
where $\widetilde{\mathcal{Q}}$ is the sample-space variable. The results are shown for $R_\lambda = 398$, $Fr=\infty$ and $Fr=0.052$, and for different $St$ and $\ell_F/\eta$. The results show, as expected, that the role of $St$ is different at different scales, which is because the behavior of $\mathcal{P}(\widetilde{\mathcal{Q}},t)$ at any scale depends upon $St_\ell$, not $St$. The results also show the strong effect of gravity on the preferential sampling, which is to suppress it. For example, in figure \ref{PDF_Q} (c), corresponding to the no gravity case, we see that preferential sampling is strongest for $St=1$. However, the results in in figure \ref{PDF_Q} (d) show that when gravity is active, the preferential sampling for $St=1$ is very weak. The suppression of preferential sampling due to gravity at any scale is because as $Fr$ is decreased, the particles fall through the flow faster, which in turn reduces the interaction time between the particles and flow eddies, thereby causing the centrifuging mechanism to be less efficient. We emphasize, however, that this does not mean that their clustering is diminished by gravity. Indeed it has been shown using DNS that for $St\gtrsim 1$, clustering is actually enhanced by gravity \citep{bec14b,ireland16b}. This is a reflection of the distinction between clustering and preferential concentration and the mechanisms responsible for each, as discussed in \S\ref{TFAS}. The subtle, but important point is that it is preferential concentration/sampling that determines the enhanced particle settling due to turbulence, and not clustering per se (see \S\ref{TFAS}). 

\begin{figure}
	\centering
	\vspace{0mm}
	{\includegraphics{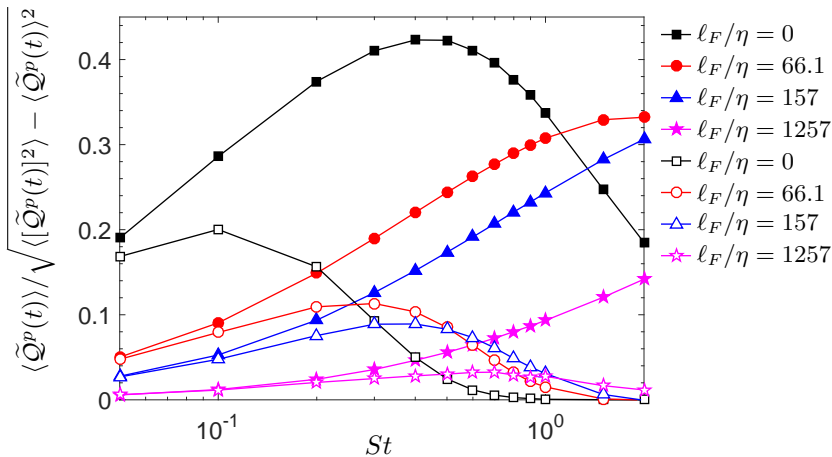}}
	\caption{DNS data for $\langle\widetilde{\mathcal{Q}}^p(t)\rangle/\sqrt{\langle[\widetilde{\mathcal{Q}}^p(t)]^2\rangle-\langle\widetilde{\mathcal{Q}}^p(t)\rangle^2}$ as a function of $St$ for different $\ell_F/\eta$, for $Fr=\infty$ (filled symbols) and $Fr=0.052$ (open symbols), and $R_\lambda=398$.}
	\label{Qnorm}
\end{figure}

Further quantitative information concerning the preferential sampling may be obtained by considering the quantity\[
\langle\widetilde{\mathcal{Q}}^p(t)\rangle/\sqrt{\langle[\widetilde{\mathcal{Q}}^p(t)]^2\rangle-\langle\widetilde{\mathcal{Q}}^p(t)\rangle^2}.
\]For homogeneous turbulence, this quantity is zero when measured along the trajectories of particles that do not preferentially sample the flow. The results in figure \ref{Qnorm} show that with or without gravity, the maximum value for this quantity weakens as $\ell_F$ is increased. This then implies that the maximum preferential sampling decreases with increasing scale. As $\ell_F$ is increased, we also see that the peak value of the curve shifts to larger $St$. This can be explained by noting that we would expect the preferential sampling at any scale to be maximum for $St_\ell=O(1)$, and as $\ell$ is increased, the value of $St$ for which $St_\ell = O(1)$ moves to larger $St$. 

The results in figure \ref{Qnorm} show that gravity significantly suppresses the preferential sampling at all scales, and the peak of the curves occurs at much lower $St$ than in the no gravity case. This latter point can be understood by the fact that since the eddy turnover timescale seen by the particle $T_\ell$ decreases with decreasing $Fr$ (for fixed $St$), then in order to observe $\tau_p/T_\ell=O(1)$ (at which one would expect the strongest preferential sampling) one has to go to smaller $\tau_p$ than in the case without gravity.

\begin{figure}
	\centering
	\vspace{0mm}
	{\includegraphics{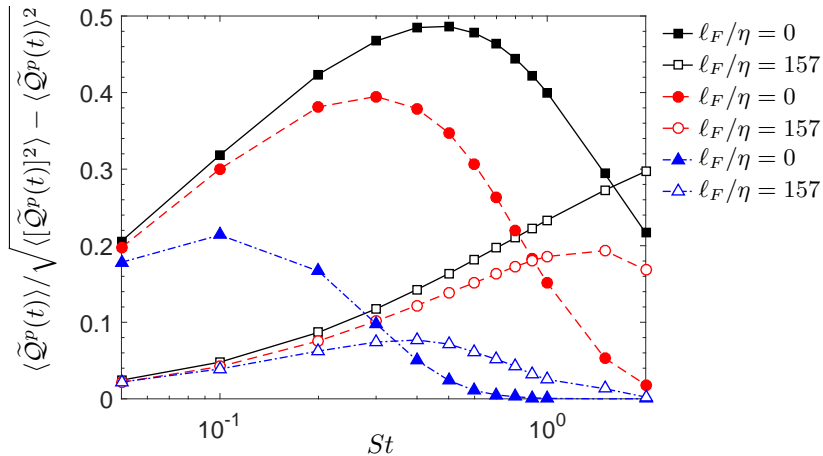}}
	\caption{DNS data for $\langle\widetilde{\mathcal{Q}}^p(t)\rangle/\sqrt{\langle[\widetilde{\mathcal{Q}}^p(t)]^2\rangle-\langle\widetilde{\mathcal{Q}}^p(t)\rangle^2}$ as a function of $St$ for $\ell_F/\eta=0$ (filled symbols) and $\ell_F/\eta=157$ (open symbols) at $R_\lambda\approx 230$, and for $Fr=\infty$ (black solid lines), $Fr=0.3$ (red dashed lines), and $Fr=0.052$ (blue dash-dot lines).}
	\label{Qnorm2}
\end{figure}

In figure \ref{Qnorm2}, we show results corresponding to figure \ref{Qnorm} but now for $R_\lambda\approx 230$ and for three different values of $Fr$ in order to further check the trends based on $Fr$ observed in figure \ref{Qnorm}. The results confirm the trends observed in figure \ref{Qnorm}, showing that as $Fr$ is decreased, the preferential sampling is systematically suppressed, and the $St$ value at which the preferential sampling is strongest shifts to smaller values.

\FloatBarrier
Finally, we pointed out earlier that results in \cite{ireland16b} showed that $\langle u_z(\bm{x}^p(t),t)\rangle>0$ may be observed for $St\geq O(1)$ and $Fr\ll 1$, even though for $St\geq O(1)$ and $Fr\ll 1$ the particles do not preferentially sample the fluid velocity gradient field. The results in figures \ref{Qnorm} and \ref{Qnorm2} for $\ell_F/\eta=0$ indeed show that for $St\geq O(1)$ and $Fr=0.052$, the particles do not preferentially sample the fluid velocity gradient field. Nevertheless, the results in figure \ref{Fig_04} show that for the same $St$ and $Fr$, $\langle u_z(\bm{x}^p(t),t)\rangle$ is significantly positive. However, the results in figures \ref{Qnorm} and \ref{Qnorm2} for $\ell_F/\eta>0$ resolve this issue because they show that while particles with $St\geq O(1)$ and $Fr=0.052$ do not preferentially sample the fluid velocity gradient field, they do preferentially sample the coarse-grained fluid velocity gradient field, i.e. $\langle\widetilde{\mathcal{Q}}^p(t)\rangle/\sqrt{\langle[\widetilde{\mathcal{Q}}^p(t)]^2\rangle-\langle\widetilde{\mathcal{Q}}^p(t)\rangle^2}$ becomes finite for $St\geq O(1)$ and $Fr=0.052$ as $\ell_F/\eta$ is increased. This confirms the picture presented by our theoretical analysis that as $St$ is increased, the scales responsible for the enhanced particle settling via preferential sweeping become larger.

\section{Conclusions}\label{sec:Conclusions}

In this paper, we have considered the multiscale nature of the mechanism leading to the enhanced settling speeds of small, heavy particles in isotropic turbulence. The traditional explanation of \citet{maxey87} is that enhanced particle settling in turbulence occurs because inertial particles preferentially sample the fluid velocity gradient field, exhibiting a tendency to be swept around the downward moving side of vortices. This mechanism is known as the preferential sweeping mechanism \citep{wang93}. However, recent results have raised questions about the completeness of this explanation. Moreover, there are several outstanding questions concerning the role of different turbulent flow scales on the enhanced settling, and the role of the Taylor Reynolds number $R_\lambda$. The theoretical work of \citet{maxey87} is not able to answer these questions due to its restriction to particles with Stokes numbers $St\ll1$. 

To address these questions we have developed a new theoretical framework for analyzing the problem that is valid for arbitrary $St$. The theory utilizes a decomposition of the ensemble averaging operator for the system that allows us to construct a result that involves a particle velocity field that is well defined for all $St$, unlike the particle velocity field of \cite{maxey87} that is only valid for $St\ll1$. Coarse-graining decompositions are also used in the theory in order to provide insight into the role of different scales in the turbulent flow on the particle settling speeds. Our theoretical result shows that the particle settling speeds are only affected by scales of the flow with size $\ell<\ell_c(St)$, where $\ell_c(St)$ is the lengthscale beyond which the effects of particle inertia are asymptotically small. Since $\ell_c(St)$ is a non-decreasing function of $St$, our theory shows shows that as $St$ is increased, increasingly larger scales contribute to the enhanced particle settling due to the turbulence. In other words, the preferential sweeping mechanism operates on progressively larger scales as $St$ is increased. Several new insights and predictions follow from our theoretical analysis, which were then tested using DNS data.

First, our theoretical analysis predicts that the range of scales contributing to the enhanced particle settling depends upon $St$, and that as a result, there can be no single turbulent flow scale that characterizes the enhanced settling. This is contrary to several previous experimental and numerical works that claim on the basis of their data that the enhanced settling speeds depend on the r.m.s velocity of the turbulence $u'$ (associated with the large scales of the flow). Therefore, even though previous studies have pointed to certain aspects of the multiscale nature of the problem (e.g. that the settling speeds depend on $u'$ and $\tau_\eta$), they concluded that the relevant turbulent velocity scale determining the particle settling speed is the same for any $St$. However, according to our analysis, the fluid velocity scale of the turbulence that dominates the settling enhancement depend essentially on $St$. The DNS results confirmed this prediction, showing that as $St$ is increased, progressively larger scales of the flow contribute to the enhanced settling. However, while it is true that only scales with size $\ell<\ell_c(St)$ contribute, our estimates show that $\ell_c(St)$ is larger than might be expected, such that even for $St=O(0.1)$, scales much larger than the Kolmogorov length scale $\eta$ contribute to the enhanced settling.

Second, our theoretical analysis predicts that the settling velocity of the particle will only be influenced by $R_\lambda$ when the integral length scale of the flow $L$, is smaller than $\ell_c(St)$. Once $L>\ell_c(St)$, the $R_\lambda$ dependence saturates because the particles are not affected by the additional scales of the flow that are introduced by increasing $R_\lambda$. When $St\ll 1$, $\ell_c(St)$ is relatively small and so the particle settling speed should show a weak dependence on $R_\lambda$. However, when $St=O(1)$, $\ell_c(St)$ can be large enough for the particles to feel the effects of the additional flow scales introduced by increasing $R_\lambda$. Our DNS results confirmed this prediction, and also provided evidence consistent with the idea that for any $St$, the particle settling speeds become independent of $R_\lambda$ for sufficiently large $R_\lambda$. Other DNS results also confirmed that the dominant effect of $R_\lambda$ on the particle settling speeds is through the scale separation in the flow that increases with increasing $R_\lambda$, rather than effects of intermittency. However, we did observe evidence of effects of intermittency on the settling speeds for $St\lesssim 0.3$.

Third, our theoretical analysis predicts that for a given $St$ and $R_\lambda$, as the Froude number $Fr$ is decreased, $\ell_c(St)$ increases, such that the faster the particles settle, the larger the scales that contribute to their enhanced settling. This is essentially a consequence of the fact that settling reduces the correlation timescale of the flow seen by the particles. Our DNS results confirmed this picture except for $St\lesssim 0.1$, where the opposite behavior was observed in some cases. We are unsure as to the explanation for this. 

Finally, we used our DNS data to examine the preferential sampling of the flow by the particles at different scales. The preferential sampling of the flow is part of the preferential sweeping mechanism, and our analysis suggests that the preferential sampling of the flow should occur at different scales depending on $St$ and $Fr$. To examine this we computed the statistics of the difference between the second invariants of the coarse-grained (at scale $\ell_F$) strain-rate and rotation-rate tensors evaluated at the positions of the inertial particles $\bm{x}^p(t)$. The results showed that the strongest preferential sampling at any scale $\ell_F$ is associated with increasingly larger $St$ as $\ell_F$ is increased. Moreover, for a given $St$, there is an optimum range of scales where the preferential sampling is strongest. As $Fr$ is decreased, the preferential sampling is suppressed, which is again due to the fact that settling reduces the correlation timescale of the flow seen by the particles, reducing the ability of local flow structures to modify the spatial distribution of the particles. When $St=O(1)$ and $Fr\ll1$, the particles sample the fluid velocity gradient field uniformly, yet they exhibit enhanced settling speeds due to the turbulence. This observation, which appears to contradict the traditional preferential sweeping mechanism, is explained by our theory as being due to the fact that for $St=O(1)$ and $Fr\ll1$, the scales at which the preferential sweeping mechanism operate do not lie in the dissipation range, but at larger scales. The DNS results confirm this since they show that while particles with $St=O(1)$ and $Fr\ll1$ do not preferentially sample the fluid velocity gradient field, they do preferentially sample the fluid velocity gradient field coarse-grained at scale $\ell_F$ outside the dissipation range.


\bigskip
\medskip
The authors wish to thank Mohammadreza Momenifar for providing some of the data used in this paper, as well as routines for producing some of the plots. This work used the Extreme Science and Engineering Discovery Environment (XSEDE), which is supported by National Science Foundation grant number ACI-1548562 \citep{xsede}. Specifically, the Comet cluster was used under allocation CTS170009.

\bibliographystyle{jfm}
\bibliography{jfm-paper}

\begin{thebibliography}{56}
\expandafter\ifx\csname natexlab\endcsname\relax\def\natexlab#1{#1}\fi
\def\au#1{#1} \def\ed#1{#1} \def\yr#1{#1}\def\at#1{#1}\def\jt#1{\textit{#1}}
  \def\bt#1{#1}\def\bvol#1{\textbf{#1}} \def\vol#1{#1} \def\pg#1{#1}
  \def\publ#1{#1}\def\arxiv#1{#1}\def\org#1{#1}\def\st#1{\textit{#1}}

\bibitem[Aliseda {\em et~al.\/}(2002)Aliseda, Cartellier, Hainaux \&
  Lasheras]{aliseda02}
{\sc \au{Aliseda, A.}, \au{Cartellier, A.}, \au{Hainaux, F.} \& \au{Lasheras,
  J.~C.}} \yr{2002}  \at{Effect of preferential concentration on the settling
  velocity of heavy particles in homogeneous isotropic turbulence}.  \jt{J.
  Fluid Mech.}  \bvol{468},  \pg{77--105}.

\bibitem[Ayyalasomayajula {\em et~al.\/}(2008)Ayyalasomayajula, Warhaft \&
  Collins]{sathya08a}
{\sc \au{Ayyalasomayajula, S.}, \au{Warhaft, Z.} \& \au{Collins, L.~R.}}
  \yr{2008}  \at{Modeling inertial particle acceleration statistics in
  isotropic turbulence}.  \jt{Phys. Fluids}  \bvol{20},  \pg{094104}.

\bibitem[Baker {\em et~al.\/}(2017)Baker, Frankel, Mani \& Coletti]{baker17}
{\sc \au{Baker, Lucia}, \au{Frankel, Ari}, \au{Mani, Ali} \& \au{Coletti,
  Filippo}} \yr{2017}  \at{Coherent clusters of inertial particles in
  homogeneous turbulence}.  \jt{Journal of Fluid Mechanics}  \bvol{833},
  \pg{364–398}.

\bibitem[Batchelor(1967)]{batchelor67}
{\sc \au{Batchelor, G.~K.}} \yr{1967} {\em An Introduction to Fluid
  Dynamics\/}.  \publ{Cambridge: Cambridge University Press}.

\bibitem[Bec {\em et~al.\/}(2006)Bec, Biferale, Boffetta, Celani, Cencini,
  Lanotte, Musacchio \& Toschi]{bec06a}
{\sc \au{Bec, J.}, \au{Biferale, L.}, \au{Boffetta, G.}, \au{Celani, A.},
  \au{Cencini, M.}, \au{Lanotte, A.~S.}, \au{Musacchio, S.} \& \au{Toschi, F.}}
  \yr{2006}  \at{Acceleration statistics of heavy particles in turbulence}.
  \jt{J. Fluid Mech.}  \bvol{550},  \pg{349--358}.

\bibitem[Bec {\em et~al.\/}(2014)Bec, Homann \& Ray]{bec14b}
{\sc \au{Bec, J\'er\'emie}, \au{Homann, Holger} \& \au{Ray, Samriddhi~Sankar}}
  \yr{2014}  \at{Gravity-driven enhancement of heavy particle clustering in
  turbulent flow}.  \jt{Phys. Rev. Lett.}  \bvol{112},  \pg{184501}.

\bibitem[Bragg \& Collins(2014)]{bragg14b}
{\sc \au{Bragg, A.D.} \& \au{Collins, L.R.}} \yr{2014}  \at{New insights from
  comparing statistical theories for inertial particles in turbulence: {I}.
  spatial distribution of particles.}  \jt{New J. Phys.}  \bvol{16},
  \pg{055013}.

\bibitem[Bragg {\em et~al.\/}(2012{\natexlab{{\em a\/}}})Bragg, Swailes \&
  Skartlien]{bragg12b}
{\sc \au{Bragg, A.}, \au{Swailes, D.~C.} \& \au{Skartlien, R.}}
  \yr{2012{\natexlab{{\em a\/}}}}  \at{Drift-free kinetic equations for
  turbulent dispersion}.  \jt{Phys. Rev. E}  \bvol{86},  \pg{056306}.

\bibitem[Bragg {\em et~al.\/}(2012{\natexlab{{\em b\/}}})Bragg, Swailes \&
  Skartlien]{bragg12}
{\sc \au{Bragg, A.}, \au{Swailes, D.~C.} \& \au{Skartlien, R.}}
  \yr{2012{\natexlab{{\em b\/}}}}  \at{Particle transport in a turbulent
  boundary layer: {N}on-local closures for particle dispersion tensors
  accounting for particle-wall interactions}.  \jt{Phys. Fluids}  \bvol{24},
  \pg{103304}.

\bibitem[{Bragg} {\em et~al.\/}(2015){Bragg}, {Ireland} \& {Collins}]{bragg14e}
{\sc \au{{Bragg}, A.~D.}, \au{{Ireland}, P.~J.} \& \au{{Collins}, L.~R.}}
  \yr{2015}  \at{Mechanisms for the clustering of inertial particles in the
  inertial range of isotropic turbulence}.  \jt{Phys. Rev. E}  \bvol{92},
  \pg{023029}.

\bibitem[Bragg {\em et~al.\/}(2015)Bragg, Ireland \& Collins]{bragg14d}
{\sc \au{Bragg, A.~D.}, \au{Ireland, P.~J.} \& \au{Collins, L.~R.}} \yr{2015}
  \at{On the relationship between the non-local clustering mechanism and
  preferential concentration}.  \jt{Journal of Fluid Mechanics}  \bvol{780},
  \pg{327--343}.

\bibitem[Chan \& Fung(1999)]{chan99}
{\sc \au{Chan, C~C} \& \au{Fung, J C~H}} \yr{1999}  \at{The change in settling
  velocity of inertial particles in cellular flow}.  \jt{Fluid Dynamics
  Research}  \bvol{25}~(5),  \pg{257}.

\bibitem[Clift {\em et~al.\/}(1978)Clift, Grace \& Weber]{clift78}
{\sc \au{Clift, R.}, \au{Grace, J.~R.} \& \au{Weber, M.~E.}} \yr{1978} {\em
  Bubbles, Drops, and Particles\/}.  \publ{Academic Press}.

\bibitem[Elghobashi \& Truesdell(1993)]{elghobashi93}
{\sc \au{Elghobashi, S.~E.} \& \au{Truesdell, G.~C.}} \yr{1993}  \at{On the
  two-way interaction between homogeneous turbulence and dispersed particles.
  1: Turbulence modification}.  \jt{Phys. Fluids A}  \bvol{5},
  \pg{1790--1801}.

\bibitem[Eyink \& Aluie(2009)]{eyink09}
{\sc \au{Eyink, Gregory~L.} \& \au{Aluie, Hussein}} \yr{2009}  \at{Localness of
  energy cascade in hydrodynamic turbulence. 1. smooth coarse graining}.
  \jt{Physics of Fluids}  \bvol{21}~(11),  \pg{115107}.

\bibitem[Fornari {\em et~al.\/}(2016)Fornari, Picano, Sardina \&
  Brandt]{fornari16}
{\sc \au{Fornari, Walter}, \au{Picano, Francesco}, \au{Sardina, Gaetano} \&
  \au{Brandt, Luca}} \yr{2016}  \at{Reduced particle settling speed in
  turbulence}.  \jt{Journal of Fluid Mechanics}  \bvol{808},  \pg{153–167}.

\bibitem[Fung(1993)]{fung93}
{\sc \au{Fung, J. C.~H.}} \yr{1993}  \at{Gravitational settling of particles
  and bubbles in homogeneous turbulence}.  \jt{J. Geophys. Res.}  \bvol{98},
  \pg{20287--20297}.

\bibitem[Fung(1998)]{fung98}
{\sc \au{Fung, J. C.~H}} \yr{1998}  \at{Effect of nonlinear drag on the
  settling velocity of particles in homogeneous isotropic turbulence}.
  \jt{Journal of Geophysical Research: Oceans}  \bvol{103}~(C12),
  \pg{27905--27917}.

\bibitem[Good {\em et~al.\/}(2014)Good, Ireland, Bewley, Bodenschatz, Collins
  \& Warhaft]{good14}
{\sc \au{Good, G.~H.}, \au{Ireland, P.~J.}, \au{Bewley, G.~P.},
  \au{Bodenschatz, E.}, \au{Collins, L.~R.} \& \au{Warhaft, Z.}} \yr{2014}
  \at{Settling regimes of inertial particles in isotropic turbulence}.
  \jt{Journal of Fluid Mechanics}  \bvol{759}.

\bibitem[Grabowski \& Wang(2013)]{grabowski13}
{\sc \au{Grabowski, W.~W.} \& \au{Wang, L.-P.}} \yr{2013}  \at{Growth of cloud
  droplets in a turbulent environment}.  \jt{Annu. Rev. Fluid Mech.}
  \bvol{45},  \pg{293--324}.

\bibitem[Guseva {\em et~al.\/}(2016)Guseva, Daitche, Feudel \& T\'el]{guseva16}
{\sc \au{Guseva, Ksenia}, \au{Daitche, Anton}, \au{Feudel, Ulrike} \&
  \au{T\'el, Tam\'as}} \yr{2016}  \at{History effects in the sedimentation of
  light aerosols in turbulence: The case of marine snow}.  \jt{Phys. Rev.
  Fluids}  \bvol{1},  \pg{074203}.

\bibitem[Gustavsson \& Mehlig(2011)]{gustavsson11b}
{\sc \au{Gustavsson, K.} \& \au{Mehlig, B.}} \yr{2011}  \at{Ergodic and
  non-ergodic clustering of inertial particles}.  \jt{Eur. Phys. Lett.}
  \bvol{96},  \pg{60012}.

\bibitem[Huck {\em et~al.\/}(2018)Huck, Bateson, Volk, Cartellier, Bourgoin \&
  Aliseda]{huck18}
{\sc \au{Huck, P.~D.}, \au{Bateson, C.}, \au{Volk, R.}, \au{Cartellier, A.},
  \au{Bourgoin, M.} \& \au{Aliseda, A.}} \yr{2018}  \at{The role of collective
  effects on settling velocity enhancement for inertial particles in
  turbulence}.  \jt{Journal of Fluid Mechanics}  \bvol{846},  \pg{1059–1075}.

\bibitem[Ireland {\em et~al.\/}(2016{\natexlab{{\em a\/}}})Ireland, Bragg \&
  Collins]{ireland16a}
{\sc \au{Ireland, P.J.}, \au{Bragg, A.D.} \& \au{Collins, L.R.}}
  \yr{2016{\natexlab{{\em a\/}}}}  \at{The effect of reynolds number on
  inertial particle dynamics in isotropic turbulence. part 1. simulations
  without gravitational effects}.  \jt{Journal of Fluid Mechanics}  \bvol{796},
   \pg{617--658}.

\bibitem[Ireland {\em et~al.\/}(2016{\natexlab{{\em b\/}}})Ireland, Bragg \&
  Collins]{ireland16b}
{\sc \au{Ireland, Peter~J.}, \au{Bragg, Andrew~D.} \& \au{Collins, Lance~R.}}
  \yr{2016{\natexlab{{\em b\/}}}}  \at{The effect of reynolds number on
  inertial particle dynamics in isotropic turbulence. part 2. simulations with
  gravitational effects}.  \jt{Journal of Fluid Mechanics}  \bvol{796},
  \pg{659--711}.

\bibitem[Ireland {\em et~al.\/}(2013)Ireland, Vaithianathan, Sukheswalla, Ray
  \& Collins]{ireland13}
{\sc \au{Ireland, P.~J.}, \au{Vaithianathan, T.}, \au{Sukheswalla, P.~S.},
  \au{Ray, B.} \& \au{Collins, L.~R.}} \yr{2013}  \at{Highly parallel
  particle-laden flow solver for turbulence research}.  \jt{Comput. Fluids}
  \bvol{76},  \pg{170--177}.

\bibitem[Kawanisi \& Shiozaki(2008)]{kawanisi08}
{\sc \au{Kawanisi, Kiyosi} \& \au{Shiozaki, Ryohei}} \yr{2008}  \at{Turbulent
  effects on the settling velocity of suspended sediment}.  \jt{J. Hydrol.
  Eng.}  \bvol{134},  \pg{261--266}.

\bibitem[Kiorboe(1997)]{kiorboe97}
{\sc \au{Kiorboe, T.}} \yr{1997}  \at{Small-scale turbulence, marine snow
  formation, and planktivorous feeding}.  \jt{Sci. Mar.}  \bvol{61 (Suppl. 1)},
   \pg{141--158}.

\bibitem[Maxey(1987)]{maxey87}
{\sc \au{Maxey, M.~R.}} \yr{1987}  \at{The gravitational settling of aerosol
  particles in homogeneous turbulence and random flow fields}.  \jt{J. Fluid
  Mech.}  \bvol{174},  \pg{441--465}.

\bibitem[Maxey \& Corrsin(1986)]{maxey86}
{\sc \au{Maxey, M.~R.} \& \au{Corrsin, S.}} \yr{1986}  \at{Gravitational
  settling of aerosol particles in randomly oriented cellular flow fields}.
  \jt{J. Aerosol. Sci.}  \bvol{43},  \pg{1112--1134}.

\bibitem[Maxey \& Riley(1983)]{maxey83}
{\sc \au{Maxey, M.~R.} \& \au{Riley, J.~J.}} \yr{1983}  \at{Equation of motion
  for a small rigid sphere in a nonuniform flow}.  \jt{Phys. Fluids}
  \bvol{26},  \pg{883--889}.

\bibitem[Mei(1994)]{mei94}
{\sc \au{Mei, R.}} \yr{1994}  \at{Effect of turbulence on the particle settling
  velocity in the nonlinear drag range}.  \jt{Int. J. Multiphase Flow}
  \bvol{20},  \pg{273--284}.

\bibitem[{Momenifar} {\em et~al.\/}(2018){Momenifar}, {Dhariwal} \&
  {Bragg}]{momenifar18}
{\sc \au{{Momenifar}, Mohammadreza}, \au{{Dhariwal}, Rohit} \& \au{{Bragg},
  Andrew~D.}} \yr{2018}  \at{{The influence of Reynolds and Froude number on
  the motion of settling, bidisperse inertial particles in turbulence}}.
  \jt{arXiv e-prints}  \pg{p. arXiv:1808.01537},  \arxiv{arXiv: 1808.01537}.

\bibitem[Monchaux \& Dejoan(2017)]{monchaux17}
{\sc \au{Monchaux, R.} \& \au{Dejoan, A.}} \yr{2017}  \at{Settling velocity and
  preferential concentration of heavy particles under two-way coupling effects
  in homogeneous turbulence}.  \jt{Phys. Rev. Fluids}  \bvol{2},  \pg{104302}.

\bibitem[Nemes {\em et~al.\/}(2017)Nemes, Dasari, Hong, Guala \&
  Coletti]{nemes17}
{\sc \au{Nemes, Andras}, \au{Dasari, Teja}, \au{Hong, Jiarong}, \au{Guala,
  Michele} \& \au{Coletti, Filippo}} \yr{2017}  \at{Snowflakes in the
  atmospheric surface layer: observation of particle turbulence dynamics}.
  \jt{Journal of Fluid Mechanics}  \bvol{814},  \pg{592–613}.

\bibitem[Nielsen(1984)]{nielsen84}
{\sc \au{Nielsen, Peter}} \yr{1984}  \at{On the motion of suspended sand
  particles}.  \jt{Journal of Geophysical Research: Oceans}  \bvol{89}~(C1),
  \pg{616--626}.

\bibitem[Nielsen(1993)]{nielsen93}
{\sc \au{Nielsen, Peter}} \yr{1993}  \at{Turbulence effects on the settling of
  suspended particles}.  \jt{J. Sediment. Petrol.}  \bvol{63},  \pg{835--838}.

\bibitem[Papanicolaou {\em et~al.\/}(2008)Papanicolaou, Elhakeem, Krallis,
  Prakash \& Edinger]{papanicolaou08}
{\sc \au{Papanicolaou, Athanasios (Thanos)~N}, \au{Elhakeem, Mohamed},
  \au{Krallis, George}, \au{Prakash, Shwet} \& \au{Edinger, John}} \yr{2008}
  \at{Sediment transport modeling review—current and future developments}.
  \jt{Journal of Hydraulic Engineering}  \bvol{134}~(1),  \pg{1--14}.

\bibitem[Petersen {\em et~al.\/}(2019)Petersen, Baker \& Coletti]{petersen19}
{\sc \au{Petersen, Alec~J.}, \au{Baker, Lucia} \& \au{Coletti, Filippo}}
  \yr{2019}  \at{Experimental study of inertial particles clustering and
  settling in homogeneous turbulence}.  \jt{Journal of Fluid Mechanics}
  \bvol{864},  \pg{925–970}.

\bibitem[Pope(2000)]{pope}
{\sc \au{Pope, S.~B.}} \yr{2000} {\em Turbulent Flows\/}.  \publ{New York:
  Cambridge University Press}.

\bibitem[Reeks(1977)]{reeks77}
{\sc \au{Reeks, M.~W.}} \yr{1977}  \at{On the dispersion of small particles
  suspended in an isotropic turbulent fluid}.  \jt{J. Fluid Mech.}  \bvol{83},
  \pg{529--546}.

\bibitem[Riley \& Lindborg(2012)]{riley12}
{\sc \au{Riley, James~J.} \& \au{Lindborg, Erik}} \yr{2012} {\em Recent
  Progress in Stratified Turbulence\/},  \pg{pp. 269--317}.  \publ{Cambridge
  University Press}.

\bibitem[Rosa {\em et~al.\/}(2016)Rosa, Parishani, Ayala \& Wang]{rosa16}
{\sc \au{Rosa, Bogdan}, \au{Parishani, Hossein}, \au{Ayala, Orlando} \&
  \au{Wang, Lian-Ping}} \yr{2016}  \at{Settling velocity of small inertial
  particles in homogeneous isotropic turbulence from high-resolution dns}.
  \jt{International Journal of Multiphase Flow}  \bvol{83},  \pg{217 -- 231}.

\bibitem[Rosa \& Pozorski(2017)]{rosa17}
{\sc \au{Rosa, Bogdan} \& \au{Pozorski, Jacek}} \yr{2017}  \at{Impact of
  subgrid fluid turbulence on inertial particles subject to gravity}.
  \jt{Journal of Turbulence}  \bvol{18}~(7),  \pg{634--652}.

\bibitem[Schöneborn(1975)]{schoneborn75}
{\sc \au{Schöneborn, P.-R.}} \yr{1975}  \at{The interaction between a single
  particle and an oscillating fluid}.  \jt{International Journal of Multiphase
  Flow}  \bvol{2}~(3),  \pg{307 -- 317}.

\bibitem[Shaw(2003)]{shaw03}
{\sc \au{Shaw, R.~A.}} \yr{2003}  \at{Particle-turbulence interactions in
  atmospheric clouds}.  \jt{Annu. Rev. Fluid Mech.}  \bvol{35},  \pg{183--227}.

\bibitem[Squires \& Eaton(1991)]{squires91a}
{\sc \au{Squires, K.~D.} \& \au{Eaton, J.~K.}} \yr{1991}  \at{Preferential
  concentration of particles by turbulence}.  \jt{Phys. Fluids A}  \bvol{3},
  \pg{1169--1178}.

\bibitem[Sundaram \& Collins(1997)]{sundaram97}
{\sc \au{Sundaram, S.} \& \au{Collins, L.~R.}} \yr{1997}  \at{Collision
  statistics in an isotropic, particle-laden turbulent suspension {I}. {D}irect
  numerical simulations}.  \jt{J. Fluid Mech.}  \bvol{335},  \pg{75--109}.

\bibitem[Towns {\em et~al.\/}(2014)Towns, Cockerill, Dahan, Foster, Gaither,
  Grimshaw, Hazlewood, Lathrop, Lifka, Peterson, Roskies, Scott \&
  Wilkins-Diehr]{xsede}
{\sc \au{Towns, J.}, \au{Cockerill, T.}, \au{Dahan, M.}, \au{Foster, I.},
  \au{Gaither, K.}, \au{Grimshaw, A.}, \au{Hazlewood, V.}, \au{Lathrop, S.},
  \au{Lifka, D.}, \au{Peterson, G.~D.}, \au{Roskies, R.}, \au{Scott, J.~R.} \&
  \au{Wilkins-Diehr, N.}} \yr{2014}  \at{Xsede: Accelerating scientific
  discovery}.  \jt{Computing in Science \& Engineering}  \bvol{16}~(5),
  \pg{62--74}.

\bibitem[Tunstall \& Houghton(1968)]{tunstall67}
{\sc \au{Tunstall, E.B.} \& \au{Houghton, G.}} \yr{1968}  \at{Retardation of
  falling spheres by hydrodynamic oscillations}.  \jt{Chemical Engineering
  Science}  \bvol{23}~(9),  \pg{1067 -- 1081}.

\bibitem[{van Hinsberg} {\em et~al.\/}(2012){van Hinsberg}, {Thije Boonkkamp},
  {Toschi} \& {Clercx}]{vanhinsberg12}
{\sc \au{{van Hinsberg}, M.~A.~T.}, \au{{Thije Boonkkamp}, J.~H.~M.},
  \au{{Toschi}, F.} \& \au{{Clercx}, H.~J.~H.}} \yr{2012}  \at{On the
  efficiency and accuracy of interpolation methods for spectral codes}.
  \jt{SIAM J. Sci. Comput.}  \bvol{34}~(4),  \pg{B479--B498}.

\bibitem[Wang \& Maxey(1993)]{wang93}
{\sc \au{Wang, L.~P.} \& \au{Maxey, M.~R.}} \yr{1993}  \at{Settling velocity
  and concentration distribution of heavy particles in homogeneous isotropic
  turbulence}.  \jt{J. Fluid Mech.}  \bvol{256},  \pg{27--68}.

\bibitem[Wilkinson \& Mehlig(2005)]{wilkinson05}
{\sc \au{Wilkinson, M.} \& \au{Mehlig, B.}} \yr{2005}  \at{Caustics in
  turbulent aerosols}.  \jt{Europhys. Lett.}  \bvol{71},  \pg{186--192}.

\bibitem[Yang \& Lei(1998)]{yang98}
{\sc \au{Yang, C.~Y.} \& \au{Lei, U.}} \yr{1998}  \at{The role of turbulent
  scales in the settling velocity of heavy particles in homogeneous isotropic
  turbulence}.  \jt{J. Fluid Mech.}  \bvol{20},  \pg{179--205}.

\bibitem[Yang \& Shy(2003)]{yang03}
{\sc \au{Yang, T.~S.} \& \au{Shy, S.~S.}} \yr{2003}  \at{The settling velocity
  of heavy particles in an aqueous near-isotropic turbulence}.  \jt{Phys.
  Fluids}  \bvol{15},  \pg{868--880}.

\bibitem[Zaichik \& Alipchenkov(2009)]{zaichik09}
{\sc \au{Zaichik, L.~I.} \& \au{Alipchenkov, V.~M.}} \yr{2009}  \at{Statistical
  models for predicting pair dispersion and particle clustering in isotropic
  turbulence and their applications}.  \jt{New J. Phys.}  \bvol{11},
  \pg{103018}.

\end{thebibliography}

\end{document}